\documentclass[times,twocolumn,final]{elsarticle}

\usepackage{framed,multirow}

\usepackage{amssymb}
\usepackage{latexsym}
\usepackage{dsfont}
\usepackage{amsmath} 
\usepackage{amsfonts} 

\usepackage{xcolor}
\usepackage{amssymb}
\usepackage{algorithm2e}

\usepackage{arydshln}

\newtheorem{prop}{Proposition}
\newproof{pf}{Proof}

\usepackage{url}
\usepackage{xcolor}
\usepackage{hyperref}
\definecolor{newcolor}{rgb}{.8,.349,.1}

\journal{Medical Image Analysis}

\begin{document}
	
	
	\begin{frontmatter}
		
		\title{Weakly Supervised Instance Learning for Thyroid Malignancy Prediction from Whole Slide Cytopathology Images} 

		\author[1]{David~Dov\corref{cor1}}
		\ead{david.dov@duke.edu}
		
		\cortext[cor1]{Corresponding author: }
		\author[2]{Shahar~Z.~Kovalsky}
		
		\author[1]{Serge~Assaad}
		
		\author[3]{Jonathan~Cohen}
		
		\author[4]{Danielle~Elliott~Range}
		\author[4]{Avani~A.~Pendse}
		
		\author[1]{Ricardo~Henao}
		\author[1]{Lawrence~Carin}
		
		\address[1]{Department of Electrical and Computer Engineering, Duke University, Durham, NC 27708, USA}
		\address[2]{Department of  Mathematics, Duke University, Durham, NC 27708, USA}
		\address[3]{Department of Surgery, Duke University Medical Center, Durham, NC 27710, USA}
		\address[4]{Department of Pathology, Duke University Medical Center, Durham, NC 27710, USA}
		

		\begin{abstract}
			We consider machine-learning-based thyroid-malignancy prediction from cytopathology whole-slide images (WSI). Multiple instance learning (MIL) approaches, typically used for the analysis of WSIs, divide the image (bag) into patches (instances), which are used to predict a single bag-level label. These approaches perform poorly in cytopathology slides due to a unique bag structure: sparsely located informative instances with varying characteristics of abnormality. We address these challenges by considering multiple types of labels: bag-level malignancy and ordered diagnostic scores, as well as instance-level informativeness and abnormality labels. We study their contribution beyond the MIL setting by proposing a maximum likelihood estimation (MLE) framework, from which we derive a two-stage deep-learning-based algorithm. The algorithm identifies informative instances and assigns them local malignancy scores that are incorporated into a global malignancy prediction. We derive a lower bound of the MLE, leading to an improved training strategy based on weak supervision, that we motivate through statistical analysis. The lower bound further allows us to extend the proposed algorithm to simultaneously predict multiple bag and instance-level labels from a single output of a neural network. Experimental results demonstrate that the proposed algorithm provides competitive performance compared to several competing methods, achieves (expert) human-level performance, and allows augmentation of human decisions.
		\end{abstract}
		
		
	\end{frontmatter}
	
	\section{Introduction}
	The prevalence of thyroid cancer is increasing worldwide \citep{aschebrook2013clinical}. The most important test in the \emph{preoperative} diagnosis of thyroid malignancy is the analysis of a fine needle aspiration biopsy (FNAB), which is stained and smeared onto a glass slide. The FNAB sample is examined under an optical microscope by a cytopathologist, who determines the risk of malignancy according to various features of follicular (thyroid) cells, such as their size, color and the architecture of cell groups. The diagnosis of FNAB, however, involves substantial clinical uncertainty and often results in unnecessary surgery.
	
	We consider the prediction of thyroid malignancy from FNAB, for which we have established in \citet{dov2019mlhc,elliott2020application} a dataset of $908$ samples. Each sample comprises a whole slide image (WSI) scanned at a typical resolution of $\sim40,000\times 25,000$ pixels, as well as the postoperative histopathology diagnosis, that is considered the ground truth in this study. The goal in this paper is to predict the ground truth malignancy label from the WSIs. Each sample also includes the diagnostic score assigned to the slide by a cytopathologist according to the Bethesda System (TBS) \citep{cibas2009bethesda}, which is the universally accepted reporting system for thyroid FNAB (there are six TBS categories). TBS $2$ indicates a benign slide, TBS $3$, $4$ and $5$ reflect inconclusive findings with an increased risk of malignancy, and TBS $6$ indicates malignancy. TBS $1$ is assigned to inadequately prepared slides and is out of the scope of this work. Further, we consider a set of $4494$ manually annotated \emph{local} labels of informative image regions containing follicular groups. The local labels indicate three categories of abnormality: $``0''$ - normal, $``1''$ - atypical, and $``2''$ malignant.
	
	Machine learning, and in particular deep neural networks, have become prevalent in the analysis of WSIs \citep{ozolek2014accurate, litjens2016deep, kraus2016classifying, sirinukunwattana2016locality, djuric2017precision, ilse2018attention, zhang2019pathologist, campanella2019clinical,glass2020use,glass2020use2}. Due to the large resolution of WSIs, gigabytes in size, each image is typically split into a set (bag) of small regions (instances) that are processed individually into local estimates, then aggregated into a global image-level prediction. This approach, often referred to as multiple instance learning (MIL)  \citep{quellec2017multiple}, addresses memory-capacity limitations of existing graphical processor unit (GPU) computing platforms. Widely used MIL approaches include \citet{zhang2006multiple} and \citet{kraus2016classifying}, which propose to aggregate local predictions via \emph{noisy-or} or \emph{noisy-and} pooling functions, respectively. In \citet{ilse2018attention} a weighted combination of local decisions is proposed, incorporating an attention mechanism to form a global decision.
	
\begin{figure}
		\begin{centering}
			\includegraphics[scale=0.7]{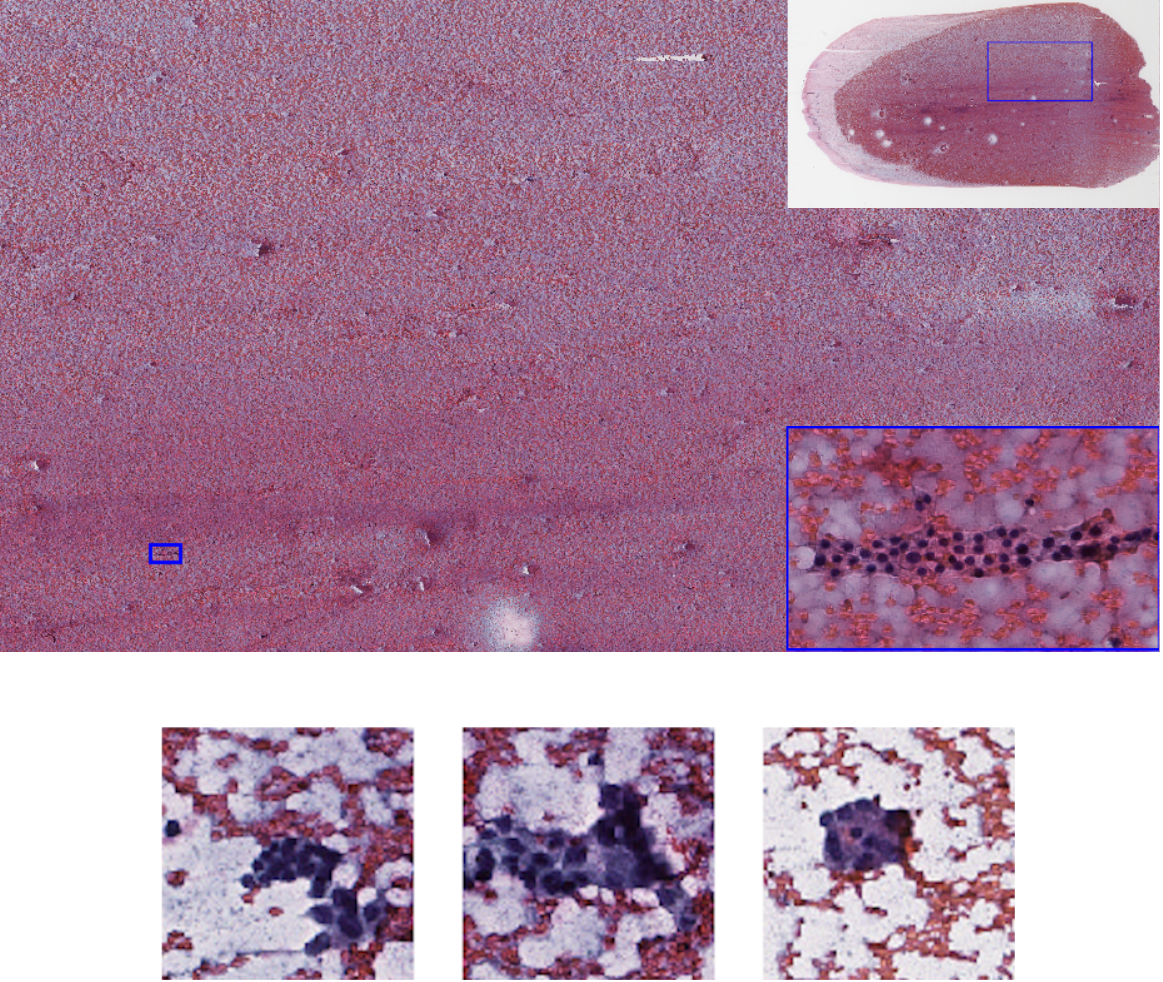}
			\par\end{centering}
		\caption{(Top) Example of a large image region of $10000\times5000$ pixels containing only a single group of follicular cells marked by the small rectangle. Top right corner: WSI with a rectangle indicating the location of the large image region. Bottom right corner: x$10$ zoomed in image of the informative follicular group. (Bottom) Examples of follicular groups with different abnormality levels. From left to right: benign, atypical and malignant.}
		\label{fig:sparsity}
	\end{figure}

	The vast majority of previous studies consider the analysis of \emph{histopathology} biopsies, which comprise whole tissues covering large regions of the WSI. In contrast, FNABs (\emph{cytopathology} biopsies), as we consider in this paper, contain separate, sparsely located groups of follicular cells, which are informative for diagnosis. The diagnosis of the FNABs, performed by a trained (cyto-)pathologist,  includes the identification of  follicular groups followed by evaluation of their characteristics.   A WSI containing even as few as six follicular groups with a size of tens of pixels, which corresponds to less than $0.01\%$ of the area of the slide, is considered sufficient for diagnosis. FNABs are considered significantly more challenging for diagnosis by pathologists due to their sparsity, and since in many cases, the characteristics of individual follicular groups are subject to subjective interpretation. An example of a large image region of $10000\times5000$ pixels containing merely a single follicular group, as well as examples of follicular groups with different abnormality levels, are presented in Fig. \ref{fig:sparsity}. Due to these challenges, the automated analysis of FNAB is addressed in the literature in a limited scale and scope. Specifically for thyroid FNAB, \citet{daskalakis2008design, varlatzidou2011cascaded, gopinath2013computer, kim2016deep, gilshtein2017computerized, savala2018artificial, sanyal2018artificial} consider manually selected individual follicular cells in extreme magnification or a small number of ``zoomed-in'' regions. However, these studies do not address the problem of intervention-free malignancy prediction from cytopathology WSIs.

	
	The paper \citet{cheplygina2019not} surveyed MIL, semi- and weakly-supervised learning approaches. These scenarios consider classification tasks with different assumptions on the availability of training labels: in MIL, only global labels are available at the bag (WSI) level, while in semi/weakly supervised setting local labels at the instance (image region) are only partially available or are noisy\citep{zhou2018brief}. \citet{cheplygina2019not} pointed out three gaps in the existing literature of medical image analysis associated with these scenarios. In the following, we address these gaps in the context of thyroid malignancy prediction. First, \citet{cheplygina2019not} claim that MIL, semi- and weakly-supervised learning are typically studied as separate problems, despite the close relation between them. Here, we investigate how only a few local, instance-level, labels can improve prediction beyond the classical MIL setting, where only a global label at the WSI/bag level is available. This is important in medical applications, where the collection of local labels requires significant manual effort, raising the question of what kind of labels to collect and what is the expertise required for their collection. For example, a non-expert could identify informative instances containing groups of follicular cells, while only a cytopathologist expert can determine the level of their abnormality (normal/atypical/malignant). In this context, we note the closely related task of region-of-interest detection, studied extensively for object detection \citep{uijlings2013selective,girshick2014rich,girshick2015fast,ren2017faster}. However, here we are not strictly concerned with the accurate estimation of bounding boxes of individual instances, a difficult challenge in the case of cytopathology, as our goal is to predict the global per-slide label.
	
	The second gap is related to the structure of the bag in MIL in terms of the \emph{prevalence of positive instances} (PPI) in a bag, which is typically not taken into account. The classical definition of MIL assumes at least one positive instance in a positive bag, while \citet{kraus2016classifying}, for example, assume a certain number of positive instances triggering a global positive label. In our context, PPI measures the fraction of the positive instances (in a positive WSI), \emph{i.e.}, those containing follicular groups with clear characteristic of malignancy. In contrast, a positive bag also contains non-malignant follicular groups, as well as uninformative instances. The uninformative instances constitute the vast majority of the scan, mainly containing red blood cells, considered in our case as background. This forms a unique bag structure of low PPI. On the other hand, once background instances are filtered out, as we propose in our approach, the bags composed of only informative instances have a high PPI structure; namely, the follicular groups are \emph{consistent} in their indication of malignancy to a certain level, which we explore in this paper.
	
	The third gap is the question of how to use multiple labels for improving classification. To this end, we consider the joint prediction of the malignancy labels, the TBS categories, and the local abnormality labels. Since both TBS categories and the local labels correspond to the increasing probability of malignancy, we consider their joint prediction using ordinal regression \citep{gutierrez2016ordinal, mccullagh1980regression, agresti2003categorical,dorado2012ordinal}. The joint prediction is motivated by the observation that the local labels, as well as TBS categories, are a consistent proxy for the probability of malignancy \citep{jing2012group,pathak2014implementation}, and so their joint prediction induces cross-regularization.
	
	This paper extends a previous conference publication \citet{dov2019mlhc}, where we presented an algorithm that provides predictions of thyroid malignancy comparable to those of cytopathology experts (we compared to three such experts). In \citet{dov2019mlhc}, we focused on a more thorough description of the clinical problem we address and provided complete details on the dataset and its acquisition. This paper focuses on the detailed derivation and the analysis of the proposed algorithm. Novel contributions, which go beyond \citet{dov2019mlhc}, include: We propose a maximum likelihood estimation (MLE) framework for classification in the mixed setting, where multiple global and local labels are available for training. While in classical MIL, informative instances are implicitly identified, the MLE framework allows explicit identification of them using the local labels, which we show to be especially useful in the low-PPI setting. We further derive a lower bound of the MLE, which corresponds to a weakly supervised training strategy, in which the global labels are propagated to the instance level and used as noisy local labels. Statistical analysis and experiments on synthetic data show that this training strategy is particularly useful for high-PPI bags obtained by filtering out the background instances. From the lower bound of the MLE, we derive the algorithm for malignancy prediction, that is based on deep-learning and comprises two stages. The algorithm identifies instances containing groups of follicular cells and incorporates local decisions based on the informative regions into the global slide-level prediction. The lower bound of the MLE further allows us to investigate the simultaneous prediction of the global malignancy and the TBS category scores, as well as the local abnormality scores. Specifically, using ordinal regression, we extend our framework to jointly predict these labels from a single output of a neural network. Extensive cross-validation experiments comparing the proposed approach to competing methods, as well as ablation experiments, demonstrate the competitive performance of the proposed algorithm. We further show that the proposed ordinal regression approach allows application of the proposed algorithm to augment cytopathologist decisions.
	
	\section{Problem formulation\label{problem_formulation}}
	Let $\mathbb{X}=\left\{X_l\right\}$ be a set of WSIs, where $X_l=\left\{ \mathbf{x}_{l,m}\right\} $ is the set of $M_l$ instances in the $l$th WSI. The $m$th instance $\mathbf{x}_{l,m}\in\mathbb{R}^{w\times h\times3}$ is a patch from an RGB digital scan, whose width and height are $w$ and $h$, respectively. Let $\mathbb{Y}=\left\{Y_l\right\}$ be the corresponding set of malignancy labels: $Y_l\in\left\{ 0,1\right\} $, where $0$ and $1$ correspond to benign and malignant cases, respectively. The goal is to predict thyroid malignancy $\hat{Y}_l$. Similar to $\mathbb{Y}$, consider the set $\mathbb{S}=\left\{S_l\right\} $, where $S_l\in\left\{2,3,4,5,6\right\}$ is the TBS category assigned to a WSI by a pathologist.
	
	We consider an additional set of \textit{local} labels $\mathbb{U} =\left\{ U_l\right\}$, where $U_l=\left\{ \mathbf{u}_{l,m}\right\} $  and $u_{l,m}\in\left\{ 0,1\right\}$.  $u_{l,m}=1$ if instance $\mathbf{x}_{l,m}$ contains a group of follicular cells, and $u_{l,m}=0$ otherwise. Our dataset includes $4494$ such informative instances, manually selected (by a trained pathologist) from $142$ WSIs. These local labels are exploited in the proposed framework for the improved identification of the informative instances. The instances containing follicular groups are further labeled according to their abnormality, forming the set $\mathbb{V}=\left\{v_{l,m}\right\}$, $v_{l,m}\in\left\{ 0,1,2\right\}$ (normal, atypical and malignant). While in the classical MIL setting, only the set of binary malignancy labels $\mathbb{Y}$ is available, we explore in this paper the contribution of the additional label sets $\mathbb{S}$, $\mathbb{U}$ and $\mathbb{V}$ for the improved prediction of thyroid malignancy.

	\section{Proposed framework for thyroid malignancy prediction\label{sec:algorithm}}
	\subsection{MLE formulation}
	\label{subsec:mle}
	Let $\mathcal{L}$ be the likelihood over the dataset given by: 
	\begin{equation}
	\mathcal{L}\triangleq P(\mathbb{X}, \mathbb{Y},\mathbb{U}) = \prod_l P(Y_l|X_l,U_l)P(U_l|X_l)P(X_l),
	\label{eq:likelihood}
	\end{equation}
	where for simplicity we only consider at this point the sets of labels $\mathbb{Y},\mathbb{U}$. We drop the right most term by assuming a uniform distribution over the WSIs, and further assume the following conditional distribution on the label $Y_l$: 
	\begin{equation}
	Y_l|X_l, U_l \sim \textup{Bernoulli}\left(\frac{1}{\tilde{M}}\sum_{m} \sigma \left( g_{\boldsymbol\theta}(\mathbf{x}_{l,m})\right) u_{l,m}\right),
	\label{eq:bernulli}
	\end{equation}
	where $g_{\boldsymbol\theta}(\mathbf{x}_{l,m})\in \mathbb{R}$ is the output of a neural network with parameters $\boldsymbol\theta$, $\sigma(\cdot)$ is the sigmoid function, and $\tilde{M} \triangleq \sum_m u_{l,m}$ (note $\tilde{M}\ll M_l$). This statistical model suggests the estimation of $Y_l$ from an average of local, instance-level estimates $g_{\boldsymbol\theta}(\mathbf{x}_{l,m})$, weighted by $u_{lm}$ according to the level of their informativeness. We note that the true labels  $u_{l,m}$ are available only for a small subset of instances. Therefore, $u_{l,m}$ in \eqref{eq:bernulli} and throughout the paper, refers to estimates of these labels unless otherwise noted. In addition, we consider $u_{l,m}$ as binary variables for simplicity, and note that our framework can be extended to continuous variables as well. We further analyze \eqref{eq:bernulli} in Subsection \ref{subsec:analysis}. Substituting  \eqref{eq:bernulli} into \eqref{eq:likelihood} leads to the following log likelihood expression, the derivation of which is presented in Appendix 1:
	\begin{align}
	\begin{aligned}
	\log\mathcal{L} = & \sum_{l} Y_{l}\log\left[\frac{1}{\tilde{M}}\sum_{m} \sigma\left(g_{\theta}(\mathbf{x}_{l,m})\right)  u_{l,m}\right] \\
	& + (1-Y_{l})\log\left[1-\frac{1}{\tilde{M}}\sum_{m} \sigma\left(g_{\theta}(\mathbf{x}_{l,m})\right)u_{l,m}\right]\\
	& + \sum_{m}\log P(u_{l,m}|\mathbf{x}_{l,m}).
	\end{aligned}
	\label{eq:bernulli_likelihood}
	\end{align}
	Maximizing the first two terms on the right hand side of \eqref{eq:bernulli_likelihood} is equivalent to minimizing the binary cross entropy (BCE) loss in the MIL setting. For example, the average pooling method is obtained by setting $u_{l,m}=\text{const}$, and the noisy-or algorithm \citep{zhang2006multiple} is obtained by setting $u_{l,m}=0$ for all instances except the one providing the highest prediction value. 
	
	In fact,  one can obtain a more general form of MIL classifier by considering a more general form of \eqref{eq:bernulli}: $Y_l|X_l, U_l \sim \textup{Bernoulli}\left(h\left(\sum_{m}g\left(\mathbf{x}_{l,m}\right)u_{l,m}\right)\right)$, where $h(\cdot)\in \mathbb{R}$ and $g(\cdot)\in \mathbb{R}^D$. This follows from \cite{zaheer2017deep,ilse2018attention}, who showed that any function invariant to the order of the instances, \emph{i.e.}, the MIL classifier in our case, can be decomposed into the form $h\left(\sum_{m}g\left(\mathbf{x}_{l,m}\right)u_{l,m}\right)$ with a particular selection of $h, g$. The attention  mechanism of \citet{ilse2018attention}, for example, explicitly identifies informative instances, $u_{l,m}$, in a data-driven manner. \citet{hou2016patch} proposed an EM-based iterative algorithm for MIL by heuristically estimating $u_{l,m}$ in the last term of \eqref{eq:bernulli_likelihood} from instance level malignancy predictions. We show in Section~\ref{experimental_results} that classical MIL algorithms, in which selection of informative instances is implicit, completely fail to predict malignancy due to the low PPI of FNABs, which mostly comprise irrelevant background instances.
	
	Equation \eqref{eq:bernulli_likelihood} is more general than the classical MIL setting, as it also allows use of the local labels to estimate the informativeness of the instances. To that end, we propose to greedily maximize \eqref{eq:bernulli_likelihood} in two steps: we use another neural network, denoted by $r_\phi(\cdot)$, trained using the last term of \eqref{eq:bernulli_likelihood} and the local labels to estimate the informativeness of instances $u_{l,m}$ (see details in Subsection \ref{subsec:identification}), and predict slide-level malignancy from the informative instances. 
	Once trained, the network for the identification of informative instances $r_\phi(\cdot)$ is applied to the WSIs, and the estimated weights $u_{l,m}$ are set to $1$ for the $\tilde{M}$ most informative instances, and zero otherwise; hence the definition $\tilde{M} \triangleq \sum_m u_{l,m}$ in \eqref{eq:bernulli} holds. We fix $\tilde{M}=1000$ instances, a value that balances the tradeoff between having a sufficient amount of training data to predict malignancy and using instances that with high probability are informative.
	
	Once the informative instances are identified, we turn to the prediction of malignancy from the first two terms in \eqref{eq:bernulli_likelihood}. Since $\sum_m{u_{l,m}/\tilde{M}}=1$, we can write:
	\begin{align}
	\begin{aligned}
	\log\mathcal{L} = & \displaystyle \sum_l Y_{l}\log\left[\sum_{m}\sigma\left(g_{\theta}(\mathbf{x}_{l,m})\right)\frac{u_{l,m}}{\tilde{M}}\right]\\
	& +(1-Y_{l})\log\left[\sum_{m}\frac{u_{l,m}}{\tilde{M}}\left(1-\sigma\left(g_{\theta}(\mathbf{x}_{l,m})\right)\right)\right]\\
	& +\sum_{m}\log P(u_{l,m}|\mathbf{x}_{l,m}).
	\end{aligned}
	\end{align}
	Using Jensen's inequality, we get the lower bound:
	\begin{align}
	\begin{aligned}
	\log\mathcal{L} \geq & \sum_{l,m} \frac{u_{l,m}}{\tilde{M}}\left[Y_{l}\log\left(\sigma\left(g_{\theta}(\mathbf{x}_{l,m})\right)\right)\right.\\
	&  +\left.(1-Y_{l})\log\left(1-\sigma\left(g_{\theta}(\mathbf{x}_{l,m})\right)\right)\right]\\
	&  +\log P(u_{l,m}|\mathbf{x}_{l,m})\\
	\triangleq & \log \mathcal{L}^\mathbb{Y} +\sum_{l,m}\log P(u_{l,m}|\mathbf{x}_{l,m}).
	\end{aligned}
	\label{eq:lower_bound}
	\end{align}
	Recalling that $u_{l,m}$ are binary, the term $-\log \mathcal{L}^\mathbb{Y}$ is the BCE loss calculated using only the informative instances. The lower bound implies the global labels $\left\{Y_l\right\}$ are assumed to hold locally, \emph{i.e.}, separately for each instance.
	We propose to train the neural network $g_{\theta}(\cdot)$ according to \eqref{eq:lower_bound}, and consider $\left\{g_{\boldsymbol{\theta}}\left(\mathbf{x}_{l,m}\right)\right\}$ as local, instance-level, predictions of thyroid malignancy, which are averaged into a global slide-level prediction:
	\begin{equation}
	f_{\boldsymbol{\theta}}\left(X_l\right)=\frac{1}{\tilde{M}}\sum_{m}g_{\boldsymbol{\theta}}\left(\mathbf{x}_{l,m}\right)u_{l,m},
	\label{eq:proposed_mil}
	\end{equation}
	where high values of $f_{\boldsymbol{\theta}}\left(X_l\right)$ correspond to high probability of malignancy. Accordingly, the predicted slide-level thyroid malignancy $\hat{Y}_l$ is given by:
	\begin{equation}
	\hat{Y}_l=\left\{ \begin{array}{cc}
	1; & \mbox{if}\,f_{\boldsymbol{\theta}}\left(X_l\right)>\beta\\
	0; & \mbox{else}
	\end{array}\right\},
	\label{eq:pred_y_pth}
	\end{equation}
	where $\beta$ is a threshold value. 
	
	\subsection{Analysis of the lower bound in the high-PPI setting}\label{subsec:analysis}

	The extent to which the assumption that the global label holds locally and separately for each instance, which stems from \eqref{eq:lower_bound}, is directly related to the bag structure. This assumption holds perfectly in the extreme case of $\text{PPI}=1$, \emph{i.e.}, that all instances are malignant in a malignant WSI and all of them are benign in a benign WSI. Yet, PPI smaller than $1$ corresponds to a weakly supervised setting where instances are paired with noisy labels. Experimental studies, such as the those presented in \citet{alpaydin2015single, rolnick2017deep}, previously reported on the robustness of neural networks to such label noise. In this subsection, we analyze the utility of the lower bound in \eqref{eq:lower_bound}, \eqref{eq:proposed_mil} and \eqref{eq:pred_y_pth} for MIL in the high PPI setting. We note that the PPI of the bags is indeed high once the uninformative labels were filtered out, as we show by the analysis of the abnormality labels $\left\{v_{l,m}\right\}$ in Section \ref{experimental_results}.    
	
	A common practice in binary classification is to predict the conditional class probability $P(Y_l=1|X_l)$. Specifically, in standard (single-instance) classification $P(Y_l=1|X_l)$ is predicted, for example, from an output of a neural network trained with BCE loss. For simplicity, we analyze $\text{logit}\left(Y_l=1|X_l\right)$, where $\text{logit}\left(\cdot\right)\triangleq\textrm{log}\left(\frac{P\left(\cdot\right)}{1-P\left(\cdot\right)}\right)$, rather than $P(Y_l=1|X_l)$. The following proposition shows that $f_{\boldsymbol{\theta}}\left(X_l\right)$ in \eqref{eq:proposed_mil} is related directly to $\text{logit}\left(Y_l=1|\mathbf{x}_{l,m}\right)$.  
	\begin{prop}\label{prop1}
		The estimate of $\text{logit}(Y_l=1|X_l)$ is given by a linear function of $f_{\boldsymbol{\theta}}\left(X_l\right)$:
		\begin{equation}
		\text{logit}(Y_l=1|X_l)=\tilde{M}f_{\boldsymbol{\theta}}\left(X_l\right)+C,
		\label{eq:proposition}
		\end{equation} where $C$ is a constant and $\tilde{M}$ is the number of the informative instances.
	\end{prop} 
	Proposition~\ref{prop1} implies that making a prediction according to \eqref{eq:pred_y_pth} by comparing $f_{\boldsymbol{\theta}}\left(X_l\right)$ to a threshold value $\beta$ is equivalent to comparing the estimated logit function to the threshold $\gamma \triangleq \tilde{M}\beta +C$. The proof is provided in Appendix 2. We further note that the logit function is directly related to the likelihood ratio test. Using Bayes rule: $\text{logit}(Y_l=1|X_l) =\log \Lambda + P(Y=0)/P(Y=1)$, where $\Lambda$ is the likelihood ratio defined as $\Lambda\triangleq P(X_l|Y_l=1)/P(X_l|Y_l=0) $. This implies that thresholding $f_{\boldsymbol{\theta}}\left(X_l\right)$ is equivalent to applying the likelihood ratio test, widely used for hypothesis testing \citep{casella2002statistical}.
	
	Proposition~\ref{prop1} provides further insight into the training strategy suggested in \eqref{eq:lower_bound}. An implicit assumption made in the proof is that $\sigma \left(g_{\boldsymbol{\theta}}\left(\mathbf{x}_{l,m}\right) \right)$ estimates the probability $P\left(Y_l=1|\mathbf{x}_{l,m}\right)$ of the slide being malignant given a single instance $\mathbf{x}_{m}$; a similar assumption is made in the derivation of the noisy-and MIL in \cite{kraus2016classifying}. Proposition~\ref{prop1} therefore implies that $f_{\boldsymbol{\theta}}\left(X_l\right)$ predicts well the likelihood ratio provided that $g_{\boldsymbol{\theta}}\left(\mathbf{x}_{l,m}\right)$ is a good estimate of $P\left(Y_l=1|\mathbf{x}_{l,m}\right)$. Equation \eqref{eq:lower_bound} indeed suggests to directly predict the global label from each instance separately. The higher the PPI is, the lower is the noise level in the the labels used to predict $P\left(Y_l=1|\mathbf{x}_{l,m}\right)$ and, according to the proposition, the better is the global prediction of $P\left(Y_l=1|X_l\right)$. This comes in contrast to \eqref{eq:bernulli_likelihood} and, specifically, to classical MIL approaches, wherein the network is optimized to predict the global label from the \emph{multiple} instances, and there is no guarantee on the quality of predictions of individual instances.
	
	\subsection{Simultaneous prediction of multiple global and local label \label{subsec:ord_reg}}
	We now consider prediction of the TBS categories $\mathbb{S}$ and the local abnormality scores $\mathbb{V}$ using the likelihood over the full dataset $P(\mathbb{X}, \mathbb{Y},\mathbb{U}, \mathbb{S}, \mathbb{V})$. To make the computation of the likelihood tractable, we assume that $P(\mathbb{Y}|\mathbb{X},\mathbb{U}), P(\mathbb{S}|\mathbb{X},\mathbb{U})$ and $P(\mathbb{V}|\mathbb{X},\mathbb{U})$ are independent. The straightforward approach under this assumption is to extend \eqref{eq:bernulli_likelihood}  and \eqref{eq:lower_bound} by adding two cross entropy loss terms to predict the labels $\mathbb{S}$ and $\mathbb{V}$, which leads to a standard multi-label scenario. However, this does not encode the strong relation between  $\mathbb{Y}$, $\mathbb{S}$ and $\mathbb{V}$, in the sense that all indicate various abnormality (malignancy) levels. We therefore propose to encode these relations into the architecture of the neural network $g_{\theta}(\cdot)$. Specifically, we take advantage of the ordinal nature of $\mathbb{S}$ and $\mathbb{V}$, where higher values of the labels indicate a higher probability of malignancy, and propose an ordinal regression framework to predict all three types of labels from a \textit{single} output of the network. In what follows, we consider for simplicity only the prediction of the global TBS category $\mathbb{S}$. Extending the framework to predict the local labels $\mathbb{V}$ is straightforward, as our lower bound formulation in \eqref{eq:lower_bound} treats local and global labels in the same manner. 
	
	Similar to \eqref{eq:pred_y_pth}, we propose to predict the TBS category by comparing the output of the network $f_{\boldsymbol{\theta}}(X_l)$ to threshold values $\beta_0 < \beta_1 < \beta_2 < \beta_3 \in \mathbb{R}$. Recall that the TBS category takes an integer value between 2 and 6, yielding:
	\begin{equation}
	\hat{S_l}=\left\{ \begin{array}{cc}
	2; & {\rm if}\,f_{\boldsymbol{\theta}}\left(X_l\right)<\beta_{0}\\
	
	n+2; & {\rm if}\,\beta_{n-1}<f_{\boldsymbol{\theta}}\left(X_l\right)<\beta_{n},\,n\in\left[1,2,3\right]\\
	6; & {\rm if}\,f_{\boldsymbol{\theta}}\left(X_l\right)>\beta_{3}
	\end{array}\right\}.
	\label{pred_y_bts}
	\end{equation}
	The proposed framework for ordinal regression is inspired by the proportional odds model, also termed the cumulative link model \citep{mccullagh1980regression, dorado2012ordinal}. The original model suggests a relationship between $f_{\boldsymbol{\theta}}\left(X_l\right)$, the threshold $\beta_n$ and the cumulative probability $P\left(S_l -2\leq n\right)$, \emph{i.e.},
	%
	\begin{equation}
	\textrm{logit}\left(S_l-2 \leq n\right) = 
	\beta_{n} - f_{\boldsymbol{\theta}}\left(X_l\right).
	\label{eq:POM}
	\end{equation}
	The proportional odds model imposes order between different TBS by linking them to $f_{\boldsymbol{\theta}}\left(X_l\right)$ so that higher values of $f_{\boldsymbol{\theta}}\left(X_l\right)$ correspond to higher TBS categories. Recalling that the logit function is a monotone mapping of a probability function into the real line, values of $f_{\boldsymbol{\theta}}\left(X_l\right)$ that are significantly smaller than $\beta_n$ correspond to high probability that the TBS category is smaller than $n+2$.
	
	We deviate from \cite{mccullagh1980regression, dorado2012ordinal} by estimating $P\left(S_l -2> n\right)$ rather than $P\left(S_l -2\leq n\right)$, which gives (derivation presented in Appendix 3):
	
	\begin{equation}
	P\left(S_l-2>n\right) = 
	\frac{1}{1+\exp\left[-\left(f_{\boldsymbol{\theta}}\left(X_l\right)-\beta_{n}\right)\right]} .
	\label{pred_Y_BTS_sigmoid}
	\end{equation}
	We note that this deviation is not necessary for the prediction of TBS, yet it allows combining the predictions of the thyroid malignancy and the TBS category in an elegant and interpretable manner. We observe that the right term in the last equation is the sigmoid function $\sigma\left(f_{\boldsymbol{\theta}}\left(X_l\right)-\beta_{n}\right)$. Accordingly, we can train the network to predict $P\left(S_l-2>l\right)$ according to:
	\begin{align}\label{eq:loss_bts}
	\begin{aligned}
	\log \mathcal{L}^{\mathbb{S}}  \triangleq  & \sum_{l,m}u_{l,m}\sum_{n=0}^{3}S^n_{l}\log\left[\sigma\left(g_{\boldsymbol{\theta}}\left(\mathbf{x}_{l,m}\right)-\beta_{n}\right)\right] \\
	& \hspace{6mm} + \left(1-S_l^n\right)\log\left[1-\sigma\left(g_{\boldsymbol{\theta}}\left(\mathbf{x}_{l,m}\right)-\beta_{n}\right)\right],
	\end{aligned}
	\end{align}
	where $S^n_{l}=\mathds{1}(S_l-2>n)$ and $\mathds{1}(\cdot)$ is the indicator function. Specifically, maximizing $\log \mathcal{L}^{\mathbb{S}}$ is equivalent to minimizing $4$ BCE loss terms with the labels $S^n_l$, $n\in\left(0, 1, 2, 3\right)$, whose explicit relation to TBS is presented in Table \ref{tab:ordinal_gr_truth} in the Appendix. The use of $g_{\boldsymbol{\theta}}\left(\mathbf{x}_{l,m}\right)$ in \eqref{eq:loss_bts}, instead of the more natural choice of $f_{\boldsymbol{\theta}} \left(X_l\right)$, is enabled by the lower bound in \eqref{eq:lower_bound}. The lower bound also allows us to extend this framework to predict the local abnormality score, which we denote by $\log \mathcal{L}^{\mathbb{V}}$, similar to \eqref{eq:loss_bts} by considering two additional thresholds, $\gamma_0, \gamma_1$ and two corresponding BCE loss terms. 
	
	For the simultaneous prediction of thyroid malignancy, TBS category and the local labels, the total loss function is given by the sum of \eqref{eq:lower_bound}, \eqref{eq:loss_bts} and $\log \mathcal{L}^{\mathbb{V}}$. We note the similarity between $\log \mathcal{L}^{\mathbb{S}}$ in \eqref{eq:loss_bts} and $\log \mathcal{L}^{\mathbb{Y}}$ in \eqref{eq:lower_bound}, a result of our choice to estimate $P\left(S_l -2> n\right)$ rather than $P\left(S_l -2\leq n\right)$ and has the following interpretation: $\log \mathcal{L}^{\mathbb{Y}}$ can be considered a special case of ordinal regression with a single fixed threshold value of $0$. The total loss function simultaneously optimizes the parameters $\boldsymbol{\theta}$ of the network $g_{\boldsymbol{\theta}}\left(\cdot\right)$ according to $7$ classification tasks, corresponding to threshold values $\{0, \beta_0,\beta_1,\beta_2,\beta_3, \gamma_0, \gamma_1\}$.
	
	The threshold values $\{\beta_0,\beta_1,\beta_2,\beta_3\}$ (and $\{\gamma_0,\gamma_1\}$) are learned along with the parameters of the networks, via stochastic gradient descent. While the training procedure does not guarantee the correct order of $\beta_0<\beta_1<\beta_2<\beta_3$ \citep{dorado2012ordinal}, we have found in our experiments that this order is indeed preserved.
	
	We note that, in some cases, the term of the loss function corresponding to the prediction of malignancy may conflict with that of the TBS category or the local label. For example, consider a malignant case $(Y_l=1)$ with TBS category 3 assigned by a pathologist. The term of the loss, in this case, which corresponds to TBS penalize high values of $f_{\boldsymbol{\theta}}\left( X_l \right)$, whereas the term corresponding to malignancy encourages them. We therefore interpret the joint estimation of TBS category, the local labels, and malignancy as a cross-regularization scheme. 
	Given two scans with the same TBS but different final pathology, the network is trained to provide higher prediction values for the malignant case. Likewise, in the case of two scans with the same pathology but different local labels, the prediction value of the scan with the higher abnormality score is expected to be higher. Thus, the network adopts properties of the Bethesda system and the abnormality scores, such that the higher the prediction value $f_{\boldsymbol{\theta}}\left( X_l \right)$ the higher is the probability of malignancy. Yet the network is not strictly restricted to the Bethesda system and the local labels, so it can learn to provide better predictions.

	\subsection{Identification of the informative instances}\label{subsec:identification}
	We predict the informativeness of the instances using a second neural network $r_\phi(\mathbf{x}_{l,m})$, optimized according to: 
	\begin{align}
	\begin{aligned}
	\log \mathcal{L}^{\mathbb{U}} \triangleq \sum_{l,m}\Big[&u_{l,m}\log(\sigma(r_\phi(\mathbf{x}_{l,m})))\\
	+& (1-u_{l,m})\log(\sigma(r_\phi(\mathbf{x}_{l,m})))\Big],
	\end{aligned}
	\end{align}
	where here $\left\{u_{l,m}\right\}$ is the set of the local labels. The term $-\log \mathcal{L}^{\mathbb{U}}$ is the standard BCE loss obtained from the last term in \eqref{eq:lower_bound}, assuming $u_{l,m}|\mathbf{x}_{l,m} \sim \textup{Bernoulli}\left(\sigma(r_\phi(\mathbf{x}_{l,m}))\right)$. Training the network requires sufficiently many  labeled examples, the collection of which was done manually by an expert pathologist through an exhaustive examination of the slides. To make the labeling effort efficient, the cytopathologist only marked positive examples of instances containing follicular groups $\left(u_{l,m}=1\right)$. We further observed in our experiments that instances sampled randomly from the whole slide mostly contain background. Therefore, to train the network $r_\phi(\mathbf{x}_{l,m})$, we assume that $u_{l,m}=0$ for all instances in the last equation except those manually identified as informative. More specifically, we propose the following design of training batches. We use batches comprising an equal number of positive and negative examples to overcome the class imbalance. As positive examples, we take follicular groups sampled uniformly at random from the set of the labeled instances, $i.e.$, for which $u_{l,m}=1$. Negative examples are obtained by sampling uniformly at random instances from the whole slide. Since in some cases informative instances can be randomly sampled and wrongly considered uninformative, the proposed training strategy can be considered weakly supervised with noisy negative labels. 
	
	To summarize, our complete log likelihood function is:
	
	\begin{align}
	\log\mathcal{L}_\text{total} \triangleq  \log \mathcal{L}^\mathbb{Y} + \log\mathcal{L}^\mathbb{S} + \log\mathcal{L}^\mathbb{V} + \log\mathcal{L}^\mathbb{U},
	\end{align}
	where $\log\mathcal{L}_\text{total}$ is the lower bound on the full log-likelihood of the probabilistic model we assume for $\mathbb{X,Y,U,S,V}$. Note that one can further weight the different likelihood components if desired. This however is not considered in the experiments, for simplicity.
	
	\section{Experiments\label{experimental_results}}
	\subsection{PPI analysis on synthetic data}
	In Subsection \ref{subsec:malignancy_prediction}, we evaluate the performance of the proposed algorithm of predicting thyroid malignancy compared to baseline MIL algorithms, considering the two settings of low PPI, when a bag comprises all instances in the WSI, and in the high PPI, after background instances were filtered out as a preprocessing step. To better understand the effect of the PPI on the performance of the different methods, we experimented with the CIFAR10 dataset \cite{krizhevsky2009learning}, designing a MIL setting where we can synthetically control the PPI. In this experiment, we consider each image as an instance and group them into bags. A bag is assigned with a positive label if at least one instance is positive and the PPI is controlled by setting the proper number of positive and negative instances in the bag. In this manner we construct multiple MIL datasets with different PPI values and evaluate the performance of the methods for each one of them. Specifically, CIFAR10 comprises natural images from $10$ classes; we assign a positive label to an instance (image) if it belongs to one of $5$ arbitrarily chosen classes and a negative label if it belongs to the other $5$ classes. Each dataset comprises $1000$ bags, with $100$ instances each. The instances are assumed independent and are sampled uniformly at random from the original dataset, with equal probability to positive and negative bags. We note that we assume independence between the instances to facilitate the simulation, an assumption which may not hold in practice as instances from the same slide may be correlated. Given an average PPI value of a dataset, we allow slight variation of the PPI in each bag by sampling, uniformly at random, bag-level PPI values in the range of $0.8-1.2$ of the average PPI. For each MIL dataset, we train each algorithm for $30$ epochs and repeat the experiment $10$ times.  
	
\begin{figure}
		\includegraphics[viewport=10bp 0bp 400bp 265bp,clip,scale=0.6]{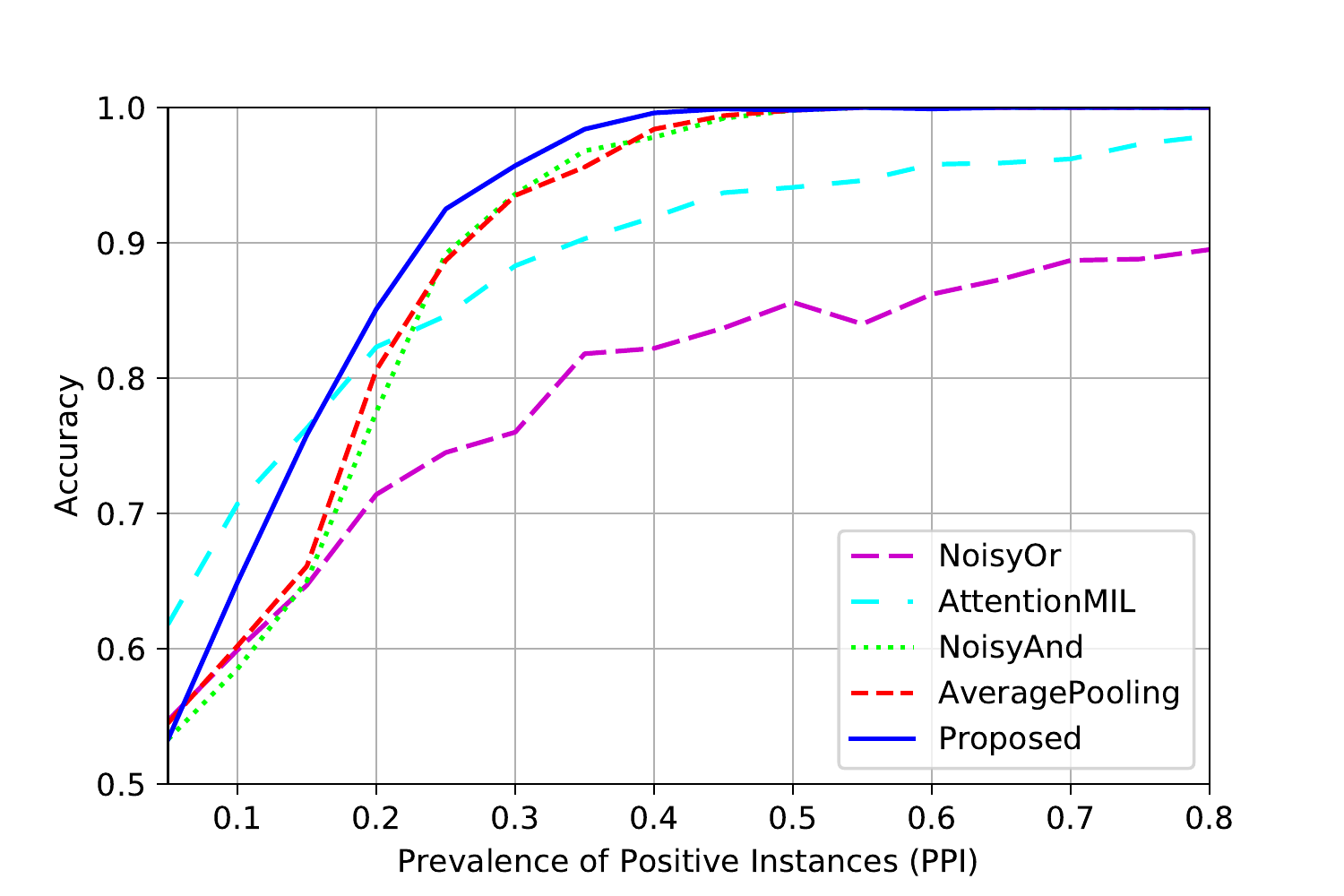}
		\caption{Accuracy \emph{vs.} PPI on CIFAR10 data.}
		\label{fig:mil_cifar}
	\end{figure}

	We compare the proposed weakly supervised training strategy derived from the lower bound in \eqref{eq:lower_bound} to the following MIL algorithms: noisy-or MIL, where the global prediction value is the highest local prediction, noisy-and MIL\ \cite{kraus2016classifying}, the attention-based MIL algorithm presented in \cite{ilse2018attention}, and average-pooling MIL obtained by maximizing the first two terms of \eqref{eq:bernulli_likelihood} rather than their lower bound. The methods are denoted ``NoisyOr,'' ``NoisyAnd,'' ``AttentionMIL" and ``AveragePooling,'' respectively. The performance of the different algorithms is presented in Fig. \ref{fig:mil_cifar}. 
	
	As expected, the performance of the methods is improved with the increase of the PPI since there are more positive instances indicating that a bag is positive. Noisy-or MIL provides inferior performance compared to the other methods for most PPI values, and only for low PPIs it performs comparably. This is because the global decision is based only on a single instance, so this approach does not benefit from the multiple positive instances present in the slides when the PPI is high. This method was excluded from the following experiments due to poor performance. 
	
	Noisy-and MIL performs on par with average-pooling, where in both methods equal weights are assigned to the different instances. The improved performance obtained by the proposed training strategy compared to average-pooling MIL supports the use of the lower bound in \ref{eq:lower_bound}, and the analysis in Section~\ref{subsec:analysis} implying that a better global prediction is obtained by training the network to directly predict the global label from each instance separately. For low PPI, the attention-based MIL provides the best performance indicating the advantage of using the attention mechanism to properly weight the instances.  The proposed training strategy performs well for high PPI values, and provides the best performance even for PPI values as low as $~0.18$. This highlights an important advantage of the proposed training strategy, that allows prediction of the global label separately from each instance, even in the presence of a large amount of label noise. 
	
	\subsection{Thyroid malignancy prediction \label{subsec:malignancy_prediction}}
	\paragraph{Experimental Setting} 
	To evaluate the proposed algorithm, we performed a $5$-fold cross-validation procedure, splitting the $908$ scans by $60\%$,  $20\%$, $20\%$ for training, validation, and testing, respectively, such that a test scan is never seen during training. The algorithm is trained using a Tesla P100-PCIE GPU with 16 Gb of memory. We use instances of size $128\times 128$ pixels. This size is large enough to capture large groups of follicular cells while allowing the use of sufficiently many instances in each minibatch. Both the network for the identification of the informative instances $r_\phi(\cdot)$ and the network for the prediction of malignancy $g_\theta(\cdot)$ are based on the small and the fast converging VGG11 architecture \cite{simonyan2014very}, details of which are summarized in Table \ref{tab:vgg11}. We observed in our experiments that the training procedure of $r_\phi(\cdot)$ converges after a few epochs, so we set a stopping criterion to avoid over-fitting. Specifically, we use the average of predictions of positive examples, a criterion we find more reasonable than, $e.g.$, the area under the (ROC) curve (AUC). The latter takes into account negative examples, the accuracy of which we are uncertain since negative examples are randomly sampled from the WSI. We stop the training process if this measure does not increase between epochs, which typically occurs after $1$ to $5$ epochs. We use 10 instances per minibatch, a value set arbitrarily and that has a small effect on the performance. The malignancy prediction network $g_\theta(\cdot)$ is trained for $100$ epochs with a minibatch size of $288$ instances, which corresponds to the maximum memory capacity of the GPU.
	\begin{figure}
		\centering
		\includegraphics[width=0.9\columnwidth]{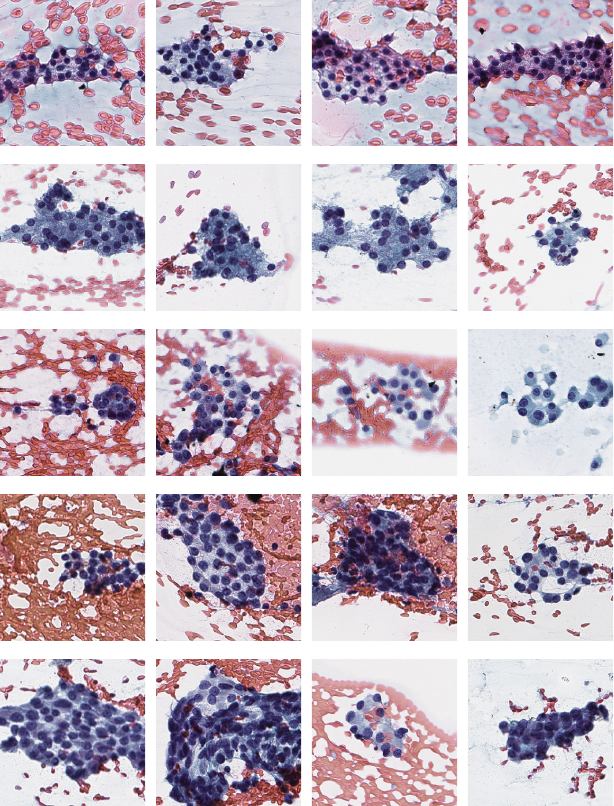}
		\caption{Instances containing follicular groups. The rows, from top to bottom, correspond to TBS $2-6$ categories.}
		\label{fig:rois_bts}
	\end{figure}
	
	%
	%
	\begin{figure}[t]
		\centering
		\includegraphics[width=.8\columnwidth]{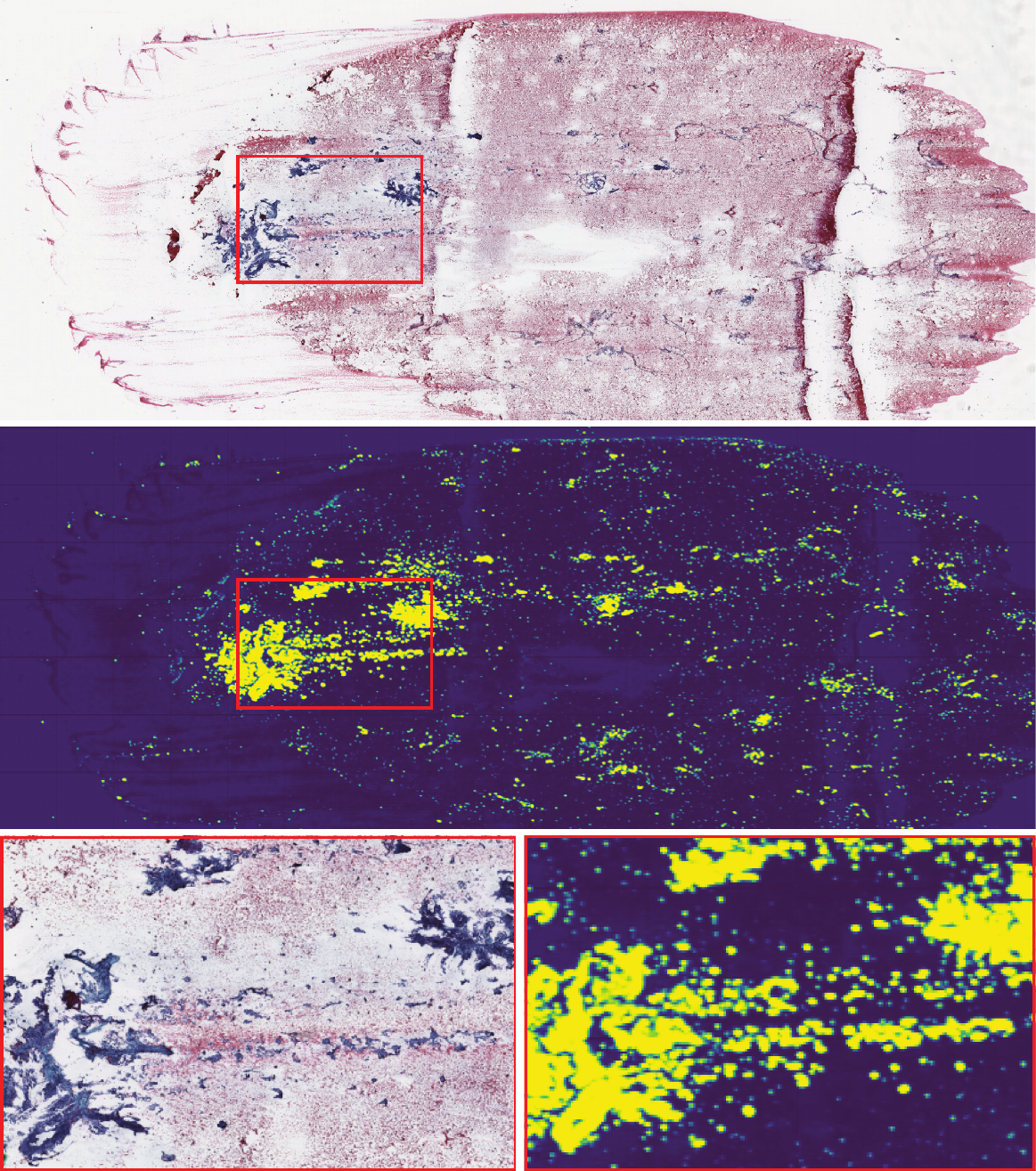}
		\caption{(Top) Whole-slide cytopathology scan. (Bottom left) Detail of the area marked by the red rectangle. (Middle) Heat map of prediction values of the first neural network. Instances predicted to contain follicular groups correspond to bright regions. (Bottom right) Detail of the are marked by the red rectangle. }
		\label{fig:full_slide_heat_map}
	\end{figure}
	\begin{figure}
		\centering
		\includegraphics[viewport=0bp 0bp 450bp 270bp,clip,scale=0.55]{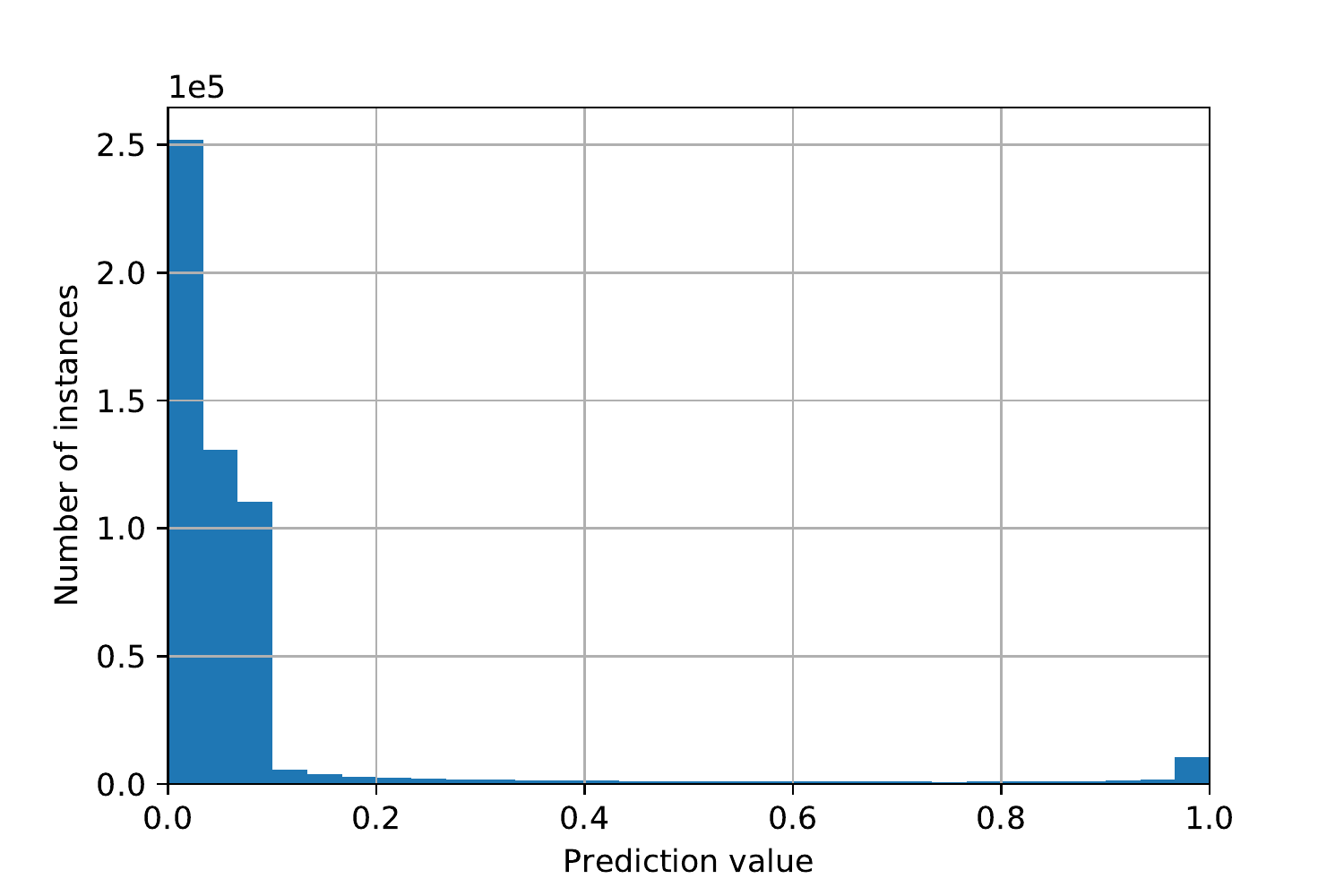}
		\caption{Histogram of predictions for instances taken from a single slide. High prediction values correspond to high probabilities that an instance contain follicular groups.}
		\label{fig:histogram_ROIS}
	\end{figure}
	
	\paragraph{Identification of instances containing follicular groups}
	
	We evaluated the performance of the network for the identification of informative instances $r_\phi(\cdot)$ using the annotated $142$ WSIs, obtaining a test AUC of $0.985$. A limitation of this analysis is that negative labels were sampled uniformly at random (as in the training procedure). We also calculated the average prediction value over the true informative instances, \emph{i.e.}, those annotated by the pathologist, and received a test average prediction value of $0.97$. Lastly, as we show below, the proposed approach leads to a significant improvement in the prediction of thyroid malignancy, in turn implying that the informative instances are indeed identified properly. In Fig.~\ref{fig:rois_bts}, we present examples of detected informative instances containing follicular groups.    
	
	We further illustrate in Figs. \ref{fig:full_slide_heat_map} and \ref{fig:histogram_ROIS} a heat map of prediction values and a corresponding histogram of informativeness predictions in an example scan. With low prediction values, the majority of the instances contain background, as is seen in both figures.
	Specifically, the follicular groups (Fig.~\ref{fig:full_slide_heat_map} top) are highlighted with bright colors in the heat map (Fig.~\ref{fig:full_slide_heat_map} middle).
	In Fig.~\ref{fig:histogram_ROIS}, the majority of instances contain background with low prediction values, however, the histogram is bimodal, with a second peak in the range of $0.95$ to $1$. These high predictions indeed correspond to instances containing follicular groups, which we select for thyroid malignancy prediction. This illustrates the extremely low PPI of FNAB WSIs, where only $\sim 2\%$ are informative and can be used to determine malignancy. We note that this example scan contains a relatively large amount of informative regions, selected for ease of presentation. In practice, the amount of informative regions can be as small as $0.01 \%$ as already stated.
	
	In this context, we note that the number, size and complexity of follicular groups are features that may indicate malignancy. Follicular group count alone is not a reliable proxy for malignancy. For example, TBS 6 slides tend to have increased numbers of large follicular groups. However, malignant slides in lower TBS categories typically have lower counts of follicular groups. Moreover, there exists benign cases (\emph{e.g.}, cases known as 'Follicular Adenomas') which exhibit similar characteristics in which the WSI is typically covered with a large number of follicular groups. For that reason, we avoid counting follicular groups.
	
	We further note that we do not consider in this paper the accurate detection of bounding boxes \cite{liu2019automated} nor pixel-level segmentation of follicular groups, rather just classifying instances of constant size as informative or not. The prediction of bounding boxes and  segmentation could allow for the explicit estimation of the size and the shape of follicular groups and have the potential to improve classification performance. Yet, these are much more challenging tasks that require a significant amount of annotation effort both for training and evaluation data. Specifically, our data set does not include accurate boxes around the bounds of the follicular groups, and in many cases, only a part of the group is annotated.
	%
	%
	\paragraph{PPI analysis}
	While the large number of background instances pose low PPI, filtering them out as a preprocessing step significantly changes the PPI in the bag. To shed light on the structure of the bag, restricted to the subset of the informative instances, we present in Table \ref{tab:local_labels} the distribution of the manually annotated local abnormality scores across the binary labels of malignancy and TBS categories. We note that the local abnormality labels were collected from an arbitrarily selected WSIs, and the cytopathologist, who was blinded to the malignancy and TBS categories of the WSIs, labeled each follicular group independently of other groups. In Table \ref{tab:local_labels} top, $80\%$ of the instances in malignant WSIs $(Y_l=1)$ are labeled  malignant, and most of them originated in TBS 6 slides. This demonstrates the  consistency between the local and the global labels, \emph{i.e.}, high PPI. Yet the PPI is lower than one: for example, the cytopathologist assigned an atypical category to $17\%$ of instances in malignant slides implying that they do not contain clear characteristics of malignancy. This demonstrates the label noise induced by the use of the lower bound in \eqref{eq:lower_bound}, according to which the global labels are propagated to the instance level. Interestingly, as seen in Table \ref{tab:local_labels} bottom, benign slides include some instances marked malignant $(v_{l,m}=2)$ by the pathologist. This contradicts the classical MIL assumption that in a negative bag all bags are negative illustrating the uncertainty in the diagnosis of cytopathology FNABs. 
	
	\begin{table}[t]
		\caption{Distribution of local abnormality labels. (Top) Distribution in malignant slides $(Y_l=1)$. (Bottom) Distribution in benign slides $(Y_l=0)$.}
		\begin{centering}
			\small
			\begin{tabular}{|c|r|r|r|}
				\cline{2-4} 
				\multicolumn{1}{c|}{} & \multicolumn{3}{|c|}{Local abnormality  labels ($v_{l,m}$)} \tabularnewline
				\hline 
				Global label & Benign $(0)$ & Atypical $(1)$ & Malignant $(2)$\tabularnewline
				\hline 
				\hline 
				$S_{l}=2$ & $15$ $(33\%)$ & $30$ $(67\%)$ & $0$ $(0\%)$\tabularnewline
				\hline 
				$S_{l}=3$ & $20$ $(38\%)$ & $32$ $(62\%)$ & $0$ $(0\%)$\tabularnewline
				\hline 
				$S_{l}=4$ & $0$ $(0\%)$ & $51$ $34\%)$ & $100$ $(66\%)$\tabularnewline
				\hline 
				$S_{l}=5$ & $4$ $(6\%)$ & $28$ $(47\%)$ & $28$ $(47\%)$\tabularnewline
				\hline 
				$S_{l}=6$ & $4$ $(0\%)$ & $110$ $(10\%)$ & $1027$ $(90\%)$\tabularnewline
				\hline 
				\hline 
				Total & $43$ $(3\%)$ & $251$ $(17\%)$ & $1155$ $(80\%)$\tabularnewline
				\hline

				\multicolumn{4}{c}{}  \tabularnewline
				
				\cline{2-4} 
				\multicolumn{1}{c|}{} & \multicolumn{3}{|c|}{Local abnormality  labels ($v_{l,m}$)} \tabularnewline
				\hline 
				Global label & Benign $(0)$ & Atypical $(1)$ & Malignant $(2)$\tabularnewline
				\hline 
				\hline 
				$S_{l}=2$ & $1234$ $(80\%)$ & $328$ $(20\%)$ & $0$ $(0\%)$\tabularnewline
				\hline 
				$S_{l}=3$ & $281$ $(40\%)$ & $373$ $(53\%)$ & $46$ $(7\%)$\tabularnewline
				\hline 
				$S_{l}=4$ & $41$ $(6\%)$ & $619$ $(89\%)$ & $38$ $(5\%)$\tabularnewline
				\hline 
				$S_{l}=5$ & $4$ $(5\%)$ & $74$ $(95\%)$ & $0$ $(0\%)$\tabularnewline
				\hline 
				$S_{l}=6$ & $0$ & $0$ & $0$\tabularnewline
				\hline 
				\hline 
				Total & $1565$ $(51\%)$ & $1396$ $(46\%)$ & $84$ $(3\%)$\tabularnewline
				\hline 
				
			\end{tabular}
			\par\end{centering}
		\hspace{2mm}
		
		\label{tab:local_labels}
	\end{table}

	\paragraph{Prediction of thyroid malignancy}
	To evaluate the proposed algorithm and the contribution of the different label sets in its design, we first consider the prediction of thyroid malignancy from the whole slide using only the global labels, and use the baseline approaches ``NoisyAND $\left(\mathbb{Y},\mathbb{S}\right)$,'' ``AttentionMIL $\left(\mathbb{Y},\mathbb{S}\right)$'' and a standard CNN (``CNN $\left(\mathbb{Y},\mathbb{S}\right)$"), all of which are trained to simultaneously predict malignancy and the TBS category (notations indicate the labels used for training). These baselines are designed originally to process whole images, which is not possible in our case due to memory limitations. Therefore, we use crops of size $448\times448$ pixels, which allows $10$ crops per minibatch, subject to memory limitations. These values were selected to optimize performance over the validation set. We compare these methods, to a version of the proposed algorithm trained according to \eqref{eq:lower_bound} without the use of the local abnormalities, denoted by ``Proposed $\left(\mathbb{U},\mathbb{Y},\mathbb{S}\right)$.'' This comparison highlights the contribution of the local label set $\mathbb{U}$ for better identification of informative instances in the low PPI setting.   
	
	
	Once we applied the network to filter out the uninformative instances, each slide is represented by a set of the informative instances only, leading to a high PPI regime. We evaluate competing MIL approached in this case also, denoting them ``NoisyAND $\left( \mathbb{U},\mathbb{Y}, \mathbb{S}\right)$,'' ``AttentionMIL $\left( \mathbb{U},\mathbb{Y}, \mathbb{S}\right)$'' and ``AveragePooling $\left( \mathbb{U},\mathbb{Y}, \mathbb{S}\right)$.  These MIL methods are trained to predict the global labels $\left( \mathbb{Y}, \mathbb{S}\right)$ from the set of the the informative instances representing each slide. We note that there is no straightforward way to incorporate the local abnormality labels $\mathbb{V}$ into the competing MIL approaches, since they are designed to use only global, slide level labels. We compare these methods to a variant of the proposed method ``Proposed $\left( \mathbb{U},\mathbb{Y},  \mathbb{S}\right)$", which uses the same labels.
	
	In addition, we consider multiple variants of the proposed algorithm, where each uses a different combination of the local and the global labels $\mathbb{Y}, \mathbb{S}$ and $\mathbb{V}$, respectively. Lastly, to better understand the advantage of using a single output of $g_\theta(\cdot)$ for the joint prediction of the labels, we consider a version termed ``Proposed2Heads $\left( \mathbb{U},\mathbb{Y}, \mathbb{S}\right)$,'' in which the network has two outputs, one for the prediction malignancy and the other for the prediction of the TBS category.
	
	\begin{table}[t]
		\caption{Comparison of the performance of the competing algorithms in the form of AUC and AP scores.}
		\begin{centering}
			\small
			\begin{tabular}{l|ll}
				\textbf{\scriptsize{}Method} & AUC & AP\tabularnewline
				\hline 
				CNN $\left(\mathbb{Y},\mathbb{S}\right)$               &  $0.748\pm0.035$ & $0.498\pm0.037$ \tabularnewline
				NoisyAND $\left(\mathbb{Y},\mathbb{S}\right)$    & $0.761\pm0.027$       & $0.538\pm0.037$\tabularnewline
				AttentionMIL $\left(\mathbb{Y},\mathbb{S}\right)$& $0.743\pm0.055$       & $0.486\pm0.095$ \tabularnewline
				\hdashline 				
				NoisyAND $\left( \mathbb{U},\mathbb{Y}, \mathbb{S}\right)$              & $0.845\pm0.016$               & $0.708\pm0.041$ \tabularnewline
				AttentionMIL $\left( \mathbb{U},\mathbb{Y},  \mathbb{S}\right)$ &  $0.823\pm0.021$        & $0.643\pm0.048$ \tabularnewline
				AveragePooling $\left(\mathbb{U},\mathbb{Y}, \mathbb{S}\right)$ & $0.850\pm0.025$ & $0.693\pm0.037$ \tabularnewline
				\hdashline 			
				Proposed $\left(\mathbb{U}, \mathbb{Y}\right)$       & $0.858\pm0.017$       & $0.719\pm0.029$ \tabularnewline
				Proposed $\left(\mathbb{U}, \mathbb{S}\right)$       & $0.852\pm0.024$       & $0.713\pm0.040$ \tabularnewline
				Proposed $\left(\mathbb{U}, \mathbb{V}\right)$       & $0.835\pm0.024$       & $0.693\pm0.049$ \tabularnewline
				Proposed $\left(\mathbb{U}, \mathbb{Y}, \mathbb{V}\right)$       & $0.858\pm0.014$       & $0.721\pm0.035$ \tabularnewline
				Proposed $\left(\mathbb{U}, \mathbb{S}, \mathbb{V}\right)$       & $0.857\pm0.018$       & $0.733\pm0.048$ \tabularnewline
				Proposed2Heads $\left(\mathbb{U}, \mathbb{Y}, \mathbb{S}\right)$  &  $0.860\pm0.019$         & $0.711\pm0.046$ \tabularnewline
				Proposed $\left(\mathbb{U}, \mathbb{Y}, \mathbb{S}\right)$  &  $\boldsymbol{0.870\pm0.017}$         & $\boldsymbol{0.743\pm0.037}$ \tabularnewline
				Proposed $\left(\mathbb{U}, \mathbb{Y}, \mathbb{S}, \mathbb{V}\right)$  &  $0.860\pm0.024$         & $0.730\pm0.047$  \tabularnewline
			\end{tabular}
			\par\end{centering}
		\hspace{2mm}
		
		\label{tab:auc_ap}
	\end{table}
	
	Table \ref{tab:auc_ap} summarizes the performance of the algorithms in the form of the area under  receiver operating characteristic curve (AUC) and the average precision (AP). As can be seen in the table,  ``CNN $(\mathbb{Y}, \mathbb{S})$," ``NoisyAND $(\mathbb{Y}, \mathbb{S})$'' and ``AttentionMIL $(\mathbb{Y, \mathbb{S}})$'' achieve markedly inferior performance compared to other methods. This is because their decisions are largely dominated by irrelevant background data. Specifically, the attention mechanism in ``AttentionMIL $(\mathbb{Y}, \mathbb{S})$'' does not properly identify the informative instances due to low PPI. The method ``Proposed $(\mathbb{U}, \mathbb{Y}, \mathbb{S})$" performs significantly better reflecting the large importance of separately identifying the informative instances according to the last term in \eqref{eq:lower_bound} using the local labels.
	
	In the high-PPI MIL setting, where each bag comprises only informative instances, ``Proposed $(\mathbb{U}, \mathbb{Y}, \mathbb{S})$'' marginally outperforms the methods ``NoisyAND $(\mathbb{U}, \mathbb{Y}, \mathbb{S})$,'' ``AttentionMIL $(\mathbb{U}, \mathbb{Y}, \mathbb{S})$'' and ``AveragePooling $(\mathbb{U}, \mathbb{Y}, \mathbb{S})$,'' and has among the lowest standard deviation. In particular, the higher AUC and AP values of the proposed algorithm, trained using the lower bound of the MLE in \eqref{eq:lower_bound}, compared to ``AveragePooling $(\mathbb{Y}, \mathbb{U}, \mathbb{S})$'' devised from \eqref{eq:bernulli_likelihood} are consistent with the experiment on synthetic data, as well as our analysis in Subsection \ref{subsec:analysis}, which suggest that better local predictions lead to improved global decisions. Moreover, as the analysis suggests, in the high-PPI setting, there is no advantage to the sophisticated aggregation of decisions from multiple instances presented in \cite{ilse2018attention}, relative to the simple averaging in \eqref{eq:proposed_mil}. 
	
	The method ``Proposed $(\mathbb{U}, \mathbb{Y}, \mathbb{S})$'' marginally outperforms all other variants of the proposed method including both ``Proposed $(\mathbb{U}, \mathbb{Y})$'' and ``Proposed2Heads $(\mathbb{U}, \mathbb{Y}, \mathbb{S})$''. This demonstrates the advantage of the proposed framework in the joint prediction of TBS categories, along with the binary malignancy labels from a single output of a neural network presented in Subsection \ref{subsec:ord_reg}. Interestingly, ``Proposed $(\mathbb{U}, \mathbb{Y}, \mathbb{S}, \mathbb{V})$'' provides inferior performance compared to ``Proposed $(\mathbb{U}, \mathbb{Y}, \mathbb{S})$''. We trained the method ``Proposed $(\mathbb{U}, \mathbb{Y}, \mathbb{S}, \mathbb{V})$'' using both the 4,494 manually annotated instances, as well as $\sim$ 545,000 instances ($\tilde M=1000$ instances per WSI), for which we considered the global labels as noisy local labels. To balance the large difference in the size of these sets and better understand the contribution of the local labels, we considered minibatches comprising of $20\%$ instances with local annotations. While it is possible to further tune the proportion of the instances with the local labels in the minibatches, this experiment suggests that there is no significant advantage to further incorporating local abnormalities scores in the proposed framework. This further demonstrates how well the network is trained using weak supervision by the global labels.
	This result further provides insight on the role the local labels and on the potential effect of inter-reviewer variability in their collection, which in our case was performed by a single pathologist. Specifically, we expect a small inter-reviewer variability in the identification of the follicular groups, which does not require a special expertise. On the other hand, assigning abnormality scores to the follicular groups can be done only by expert pathologists, and we do expect variability between reviewers. In the setting of our experiments, and under the constraints we had on collecting expert annotations, the small number of 4,494 abnormality labels did not improve the results.

	\begin{figure}
		\begin{centering}
			\includegraphics[viewport=0bp 5bp 432bp 260bp,clip,scale=0.45]{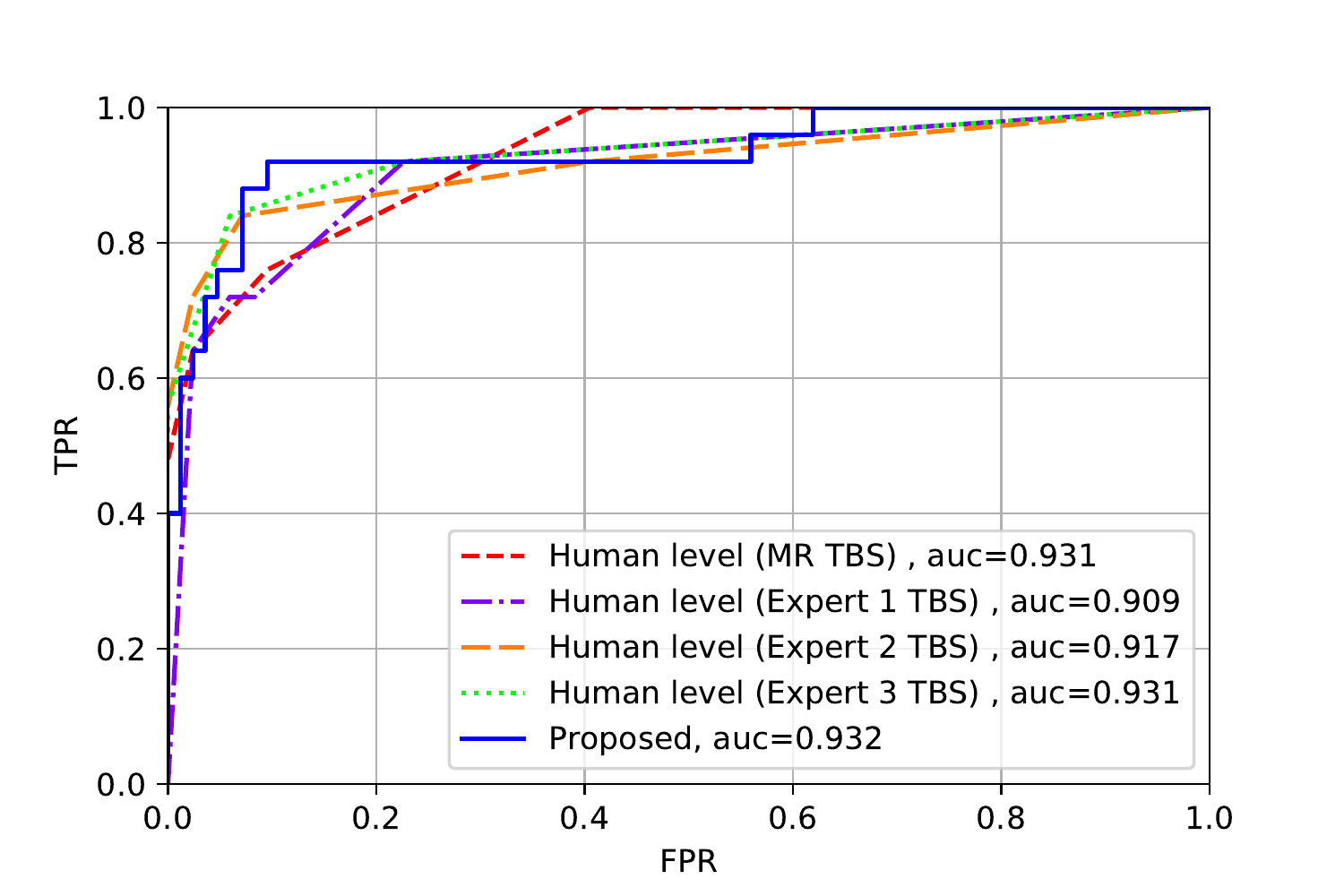}
			\includegraphics[viewport=0bp 5bp 432bp 260bp,clip,scale=0.45]{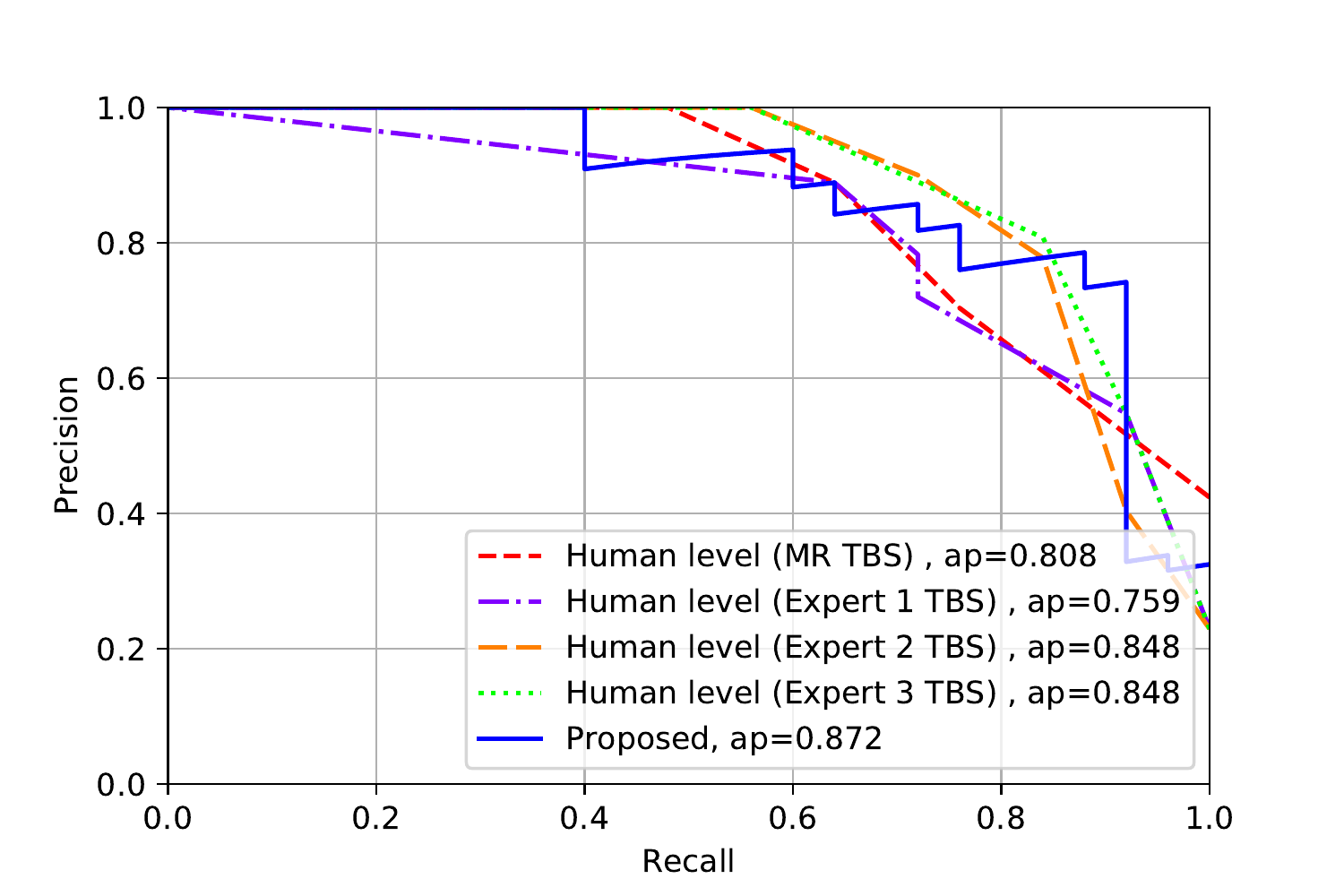}
			\par\end{centering}
		\centering{}
		\caption{ROC (Left) and PR (Right) curves comparing the performance of the proposed algorithm and human experts in predicting thyroid malignancy. Blue curve - the proposed algorithm. Red curve - pathologist from the medical record. Purple, orange  and green curves - expert cytopathologists $1$, $2$, and $3$, respectively (these three individuals analyzed the same digital image considered by the algorithm, and these experts were not the same as the clinicians from the medical record).}
		\label{fig:roc}
	\end{figure}
	
	\paragraph{Comparison to human-level performance}
	For the comparison of the algorithm to human-level performance, we use a subset of $109$ slides which were reviewed by $3$ expert cytopathologists (Experts $1$ to $3$), who assigned TBS categories, in addition to the TBS available in the original medical record (MR TBS). The performance of the proposed algorithm (``Proposed $(\mathbb{U}, \mathbb{Y}, \mathbb{S})$'') is compared to those of human in Fig.~\ref{fig:roc}, using receiver operating characteristic (ROC) and precision-recall (PR) curves. Curves representing the performance of the human experts are obtained by considering the TBS categories as ``human predictions of malignancy" such that TBS categories $2$ to $6$ correspond to increasing probability of thyroid malignancy. The AUC score obtained by the proposed algorithm is comparable to those of humans, and the algorithm provides an improved AP score compared to the human experts.
	
\begin{figure}
		\begin{centering}
			\includegraphics[viewport=0bp 0bp 280bp 280bp,clip,scale=0.6]{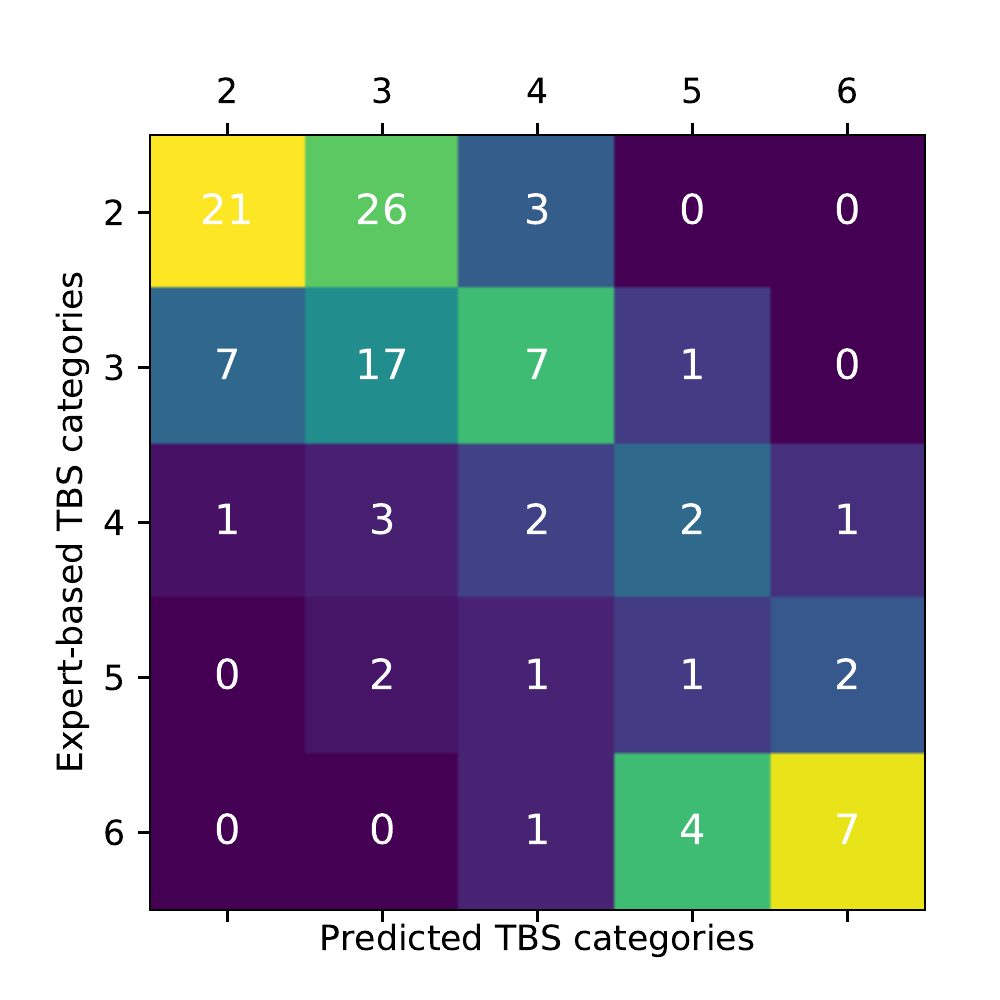}
			\par\end{centering}
		\centering{}
		\caption{Confusion matrix of TBS categories assigned by the proposed algorithm \emph{vs.} human experts. The colors in the plot correspond to a column normalized version of the confusion matrix.}
		\label{fig:tbs_vs_predicted}
	\end{figure}

	Figure~\ref{fig:tbs_vs_predicted} further presents a comparison of TBS scores assigned by the algorithm and the human experts. High values are obtained at the top-left and right-bottom of the matrix, while off-diagonal values decay. This block diagonal structure is exactly what is expected from the algorithm rather than, \emph{e.g.}, a diagonal structure. For the indeterminate cases, assigned TBS $3$ to $5$ by the experts, the term of the loss function corresponding to final pathology $\mathcal{L}^\mathbb{Y}$ encourages the algorithm to deviate from the original TBS, and provide either lower values in the benign cases or higher values in the malignant ones. On the other hand, cases assigned with TBS $2$ and $6$ by cytopathologists are benign and malignant, respectively, in more than $97\%$ of the cases. This high confidence in TBS $2$ and $6$ cases is similarly encoded in the algorithm, as we note that \emph{all} the cases for which the algorithm predicts TBS $2$ or $6$ are indeed benign or malignant, respectively. 
	
	This implies the potential to apply the algorithm for augmenting  cytopathologists' decisions. By the joint prediction of TBS and malignancy from a single output of the network, the proposed framework allows the grouping of predictions according to increasing probabilities of malignancy, using the thresholds $\{\beta_0, \beta_1, \beta_2, \beta_3\}$ in \eqref{eq:loss_bts}. This allows one to naturally combine the human and algorithm decisions according to the following rule: use human or the algorithm's decision if either of them assign TBS $2$ or $6$. In the case both of them assign an indeterminate score of TBS 3 to 5, we consider two variants: 1) use human decision, and 2) use the algorithm's decision. Table \ref{tab:combined} shows that the combined rule where indeterminate decisions are held by the algorithm indeed improves the decisions of all three experts, further implying that the algorithm performs beyond human-level in indeterminate cases.

	\begin{table}
		\caption{Combined human and algorithm decisions. Decision rule: use human or
			algorithm decisions if either of them assign TBS 2 or 6. ``Combined345Human'':
			the decision in TBS 3,4,5 cases is made by human. ``Combined345Alg:
			the decision in TBS 3,4,5 cases is by the algorithm.}
		\begin{centering}
			{\small{}}%
			\begin{tabular}{|c|c|c|c|}
				\cline{2-4} 
				\multicolumn{1}{c|}{} & {\smaller{}Human} & {\smaller{}Combined345Human} & {\smaller{}Combined345Alg}\tabularnewline
				
				\hline 
				{\small{}Expert 1} & {\small{}$0.909$} & {\small{}$0.918$} & \textbf{\small{}$\boldsymbol{0.925}$}\tabularnewline
				\hline 
				{\small{}Expert 2} & {\small{}$0.917$} & {\small{}$0.925$} & \textbf{\small{}$\boldsymbol{0.929}$}\tabularnewline
				\hline 
				{\small{}Expert 3} & {\small{}$0.931$} & {\small{}$0.934$} & \textbf{\small{}$\boldsymbol{0.937}$}\tabularnewline
				\hline 
			\end{tabular}{\small\par}
			\par\end{centering}
		{\small{}\vspace{0.3cm}
		}{\small\par}
		\begin{centering}
			{\small{}}%
			\begin{tabular}{|c|c|c|c|}
				\cline{2-4} 
				\multicolumn{1}{c|}{} & {\smaller{}Human} & {\smaller{}Combined345Human} & {\smaller{}Combined345Alg}\tabularnewline
				\hline 
				
				{\small{}Expert 1} & {\small{}$0.759$} & {\small{}$0.784$} & {\small{}$\boldsymbol{0.812}$}\tabularnewline
				\hline 
				{\small{}Expert 2} & {\small{}$0.848$} & {\small{}$0.867$} & {\small{}$\boldsymbol{0.886}$}\tabularnewline
				\hline 
				{\small{}Expert 3} & {\small{}$0.848$} & {\small{}$0.864$} & {\small{}$\boldsymbol{0.888}$}\tabularnewline
				\hline 
			\end{tabular}{\small\par}
			\par\end{centering}
		
		\label{tab:combined}
	\end{table}
	
	\section{Conclusions}
	We have considered machine-learning-based prediction of thyroid malignancy from cytopathology WSIs, in the setting where multiple local and global labels are available for training. An MLE formulation has been presented, that extends MIL to this setting, and, using a lower bound of the MLE, devised a two-stage algorithm.
	Inspired by the work of a cytopathologist, the algorithm identifies informative instance containing follicular cells, and then assigns a reliable slide-level malignancy score, similar to the Bethesda system, where higher values correspond to higher probabilities of malignancy. 
	We showed that the MLE framework facilitates the use of local labels for the improved identification of informative instances in the low-PPI regime, where most instances are uninformative. In the high-PPI setting, after the uninformative instances have been excluded, statistical analysis and experiments on both synthetic and cytopathology WSIs data showed the advantage of the weakly supervised training strategy induced by the lower bound of the MLE.  
	Experimental results further showed that the proposed framework for simultaneous prediction of binary malignancy labels and TBS categories does not benefit from the use of the manually collected abnormality scores. While a non-expert can manually identify informative instances, assigning abnormality scores requires the expertise of an expert cytopathologist and is costly and time-consuming. We showed that the proposed algorithm, without using these labels, achieves performance comparable to three cytopathologists, and demonstrated the application of the algorithm to improve human decisions. 
	The proposed MLE framework and the lower bound have two important properties that are general rather than specific to thyroid data: the framework decouples the identification and classification of instances, and it naturally associates between local and global labels.  Our future plans are to apply the framework to the diagnosis of prostate cancer, where these properties may be particularly useful. First, prostate diagnosis is determined by the classification of prostate glands, so that it may be useful first to separate them from the irrelevant background. Moreover, in prostate slides there is an explicit relation between the local and global labels: the global diagnostic score, termed the Gleason score, is given by the frequency and the severity of the local labels.


	\section*{Appendix 1}
	By substituting  \eqref{eq:bernulli} into \eqref{eq:likelihood} we get:
	\begin{equation}
	\begin{array}{ll}
	\mathcal{L} = & \prod_l \left(\frac{1}{\tilde{M}}\sum_m \left(\sigma\left(g_{\theta}(\mathbf{x}_{l,m})\right)u_{l,m}\right)\right)^{Y_l} \\
	& \cdot \left(1-\frac{1}{\tilde{M}}\sum_m(\sigma\left(g_{\theta}(\mathbf{x}_{l,m})\right)u_{l,m}\right)^{1-Y_l}\\
	& \cdot P(U_l|X_l).
	\end{array}
	\end{equation}
	We take a $\log$ from both sides of the equation:
	\begin{equation}
	\begin{array}{ll}
	\log\mathcal{L} = & \sum_l Y_l\log\left(\frac{1}{\tilde{M}}\sum_m \sigma\left(g_{\theta}(\mathbf{x}_{l,m})\right)u_{l,m}\right) \\
	& + (1-Y_l)\log\left(1-\frac{1}{\tilde{M}}\sum_m \sigma\left(g_{\theta}(\mathbf{x}_{l,m})\right)u_{l,m}\right)  \\
	& +\log P(U_l|X_l), 
	\end{array}
	\end{equation}
	and we get \eqref{eq:bernulli_likelihood} by rewriting the right term by assuming that being an instance informative is independent of other instances.
	
	\section*{Appendix 2}
	\textit{Proposition 1}
	The estimate of $\text{logit}(Y_l=1|X_l)$ is given by a linear function of $f_{\boldsymbol{\theta}}\left(X_l\right)$:
	\begin{equation}
	\text{logit}(Y_l=1|X_l)=\tilde{M}f_{\boldsymbol{\theta}}\left(X_l\right)+C,
	\label{eq:proposition_proof}
	\end{equation} where $C$ is a constant and $\tilde{M}$ is the number of the informative instances.
	
	The proof is based on the assumption that the instances $\mathbf{x}_{l,m}$ are independent random variables. We note that this assumption is used to facilitate the derivation and it might not hold in practice for instances taken from the same scan. Yet, we motivate this assumption by the large variability between the follicular groups in their size, architecture and the number of cells as demonstrated in Fig. \ref{fig:rois_bts}. 
	\begin{pf}
		From the definition of $\text{logit}(Y_l=1|X_l)$, and using the Bayes rule we get:
		\begin{align}\label{eq:log_lr}
		\begin{aligned}
		\text{logit}(Y_l=1|X_l)&=\log\left(\frac{P\left(Y_l=1|X_l\right)}{P\left(Y_l=0|X_l\right)}\right) \\
		&=\log\left(\frac{P\left(X_l|Y_l=1\right)}{P\left(X_l|Y_l=0\right)}\right) + \log\left(\frac{P\left(Y_l=0\right)}{P\left(Y_l=1\right)}\right).
		\end{aligned}
		\end{align}
		By further using the independence assumption, we have:
		\begin{equation}
		\text{logit}(Y_l=1|X_l)=\sum_{m}\log\left(\frac{P\left(\mathbf{x}_{l,m}|Y_l=1\right)}{P\left(\mathbf{x}_{l,m}|Y_l=0\right)}\right) + \log\left(\frac{P\left(Y_l=0\right)}{P\left(Y_l=1\right)}\right).
		\end{equation}
		Since for the uninformative instances  $P\left(\mathbf{x}_{l,m}|Y_l=1\right)=P\left(\mathbf{x}_{l,m}|Y_l=0\right)$, the sum in last equation is in fact over the $\tilde{M}$ informative instances rather than over the whole set of size $M$. Another application of the the Bayes on the first right term leads to:
		\begin{align}
		\begin{aligned}
		\text{logit}(Y_l=1|X_l)=&\sum_{m}\log\left(\frac{P\left(Y_l=1|\mathbf{x}_{l,m}\right)P\left(Y_l=0\right)}{P\left(Y_l=0|\mathbf{x}_{l,m}\right)P\left(Y_l=1\right)}\right) \\
		&+\log\left(\frac{P\left(Y_l=0\right)}{P\left(Y_l=1\right)}\right) \\
		=&\sum_{m}\log\left(\frac{P\left(Y_l=1|\mathbf{x}_{l,m}\right)}{P\left(Y_l=0|\mathbf{x}_{l,m}\right)}\right) + C \\
		=&\sum_{m}\text{logit}\left(Y_l=1|\mathbf{x}_{l,m}\right) + C,
		\label{eq:lrt_logit}
		\end{aligned}
		\end{align}
		where: $C\triangleq (\tilde{M}+1)\log\left(\frac{P\left(Y_l=0\right)}{P\left(Y_l=1\right)}\right)$. According to \eqref{eq:lower_bound}, the last equation is estimated by:
		\begin{equation}
		\text{logit}(Y_l=1|X_l)=\sum_{m}u_{l,m} g_{\boldsymbol{\theta}}\left(\mathbf{x}_{l,m}\right)+C .
		\label{eq:lrt_pred}
		\end{equation}
		Finally, \eqref{eq:proposition_proof} is given by assigning \eqref{eq:proposed_mil} into \eqref{eq:lrt_pred}. 
	\end{pf}
	
	\section*{Appendix 3}
	By definition of the logit function and since $P\left(S_l -2\leq n\right) = 1 - P\left(S_l-2> n\right)$ we have:
	\begin{equation}
	\textrm{logit}\left(S_l-2 > n\right) = 
	-\textrm{logit}\left(S_l-2 \leq n\right).
	\end{equation}
	Further substituting the last equation into \eqref{eq:POM}, gives:
	\begin{equation}
	\textrm{logit}\left(S_l-2 > n\right) = 
	-\left(f_{\boldsymbol{\theta}}(X_l)-\beta_{n}\right). 
	\label{eq:POM_proposed}
	\end{equation}
	Last, we rewrite (\ref{eq:POM_proposed}) as
	\begin{equation}
	P\left(S_l-2>n\right) = 
	\frac{1}{1+\exp\left[-\left(f_{\boldsymbol{\theta}}\left(X_l\right)-\beta_{n}\right)\right]} .
	\end{equation}
	%
	
	\begin{table}[t]
		\caption{VGG11 based architecture used for both the first and the second neural
			networks in the proposed algorithm. Each conv2d layer comprises 2D
			convolutions with the parameters $\text{kernel\_size}=3$ and $\textrm{padding}=1$.
			Parameters of the Max-pooling layer: $\textrm{kernel\_size}=2$, $\textrm{stride}=2$.
			The conv2d and the linear layers (except the last one) are followed
			by batch normalization and ReLU. The network is trained using the binary cross entropy (BCE) loss via stochastic gradient descent with learning rate $0.001$, momentum $0.99$ and weight decay with decay parameter $10^{-7}$.}
		\begin{centering}
			{\scriptsize{}}%
			\begin{tabular}{|c|c|}
				\hline 
				\multicolumn{2}{|c|}{\textbf{\scriptsize{}Feature extraction layers}}\tabularnewline
				\hline 
				\hline 
				\textbf{\scriptsize{}Layer} & \textbf{\scriptsize{}Number of filters}\tabularnewline
				\hline 
				{\scriptsize{}conv2d} & $64$\tabularnewline
				\hline 
				{\scriptsize{}Max-pooling(M-P)} & \tabularnewline
				\hline 
				{\scriptsize{}conv2d} & $128$\tabularnewline
				\hline 
				M-P & \tabularnewline
				\hline 
				{\scriptsize{}conv2d} & $256$\tabularnewline
				\hline 
				{\scriptsize{}conv2d} & $256$\tabularnewline
				\hline 
				M-P & \tabularnewline
				\hline 
				{\scriptsize{}conv2d} & $512$\tabularnewline
				\hline 
				{\scriptsize{}conv2d} & $512$\tabularnewline
				\hline 
				M-P & \tabularnewline
				\hline 
			\end{tabular}{\scriptsize\par}
			\par\end{centering}
		\medskip{}
		
		\begin{centering}
			\begin{tabular}{|c|c|}
				\hline 
				\multicolumn{2}{|c|}{\textbf{\scriptsize{}Classification layers}}\tabularnewline
				\hline 
				\hline 
				\textbf{\scriptsize{}Layer} & \textbf{\scriptsize{}Output size}\tabularnewline
				\hline 
				{\scriptsize{}Linear} & {\scriptsize{}$4096$}\tabularnewline
				\hline 
				{\scriptsize{}Linear} & {\scriptsize{}$4096$}\tabularnewline
				\hline 
				{\scriptsize{}Linear} & {\scriptsize{}$1$}\tabularnewline
				\hline 
			\end{tabular}{\scriptsize\par}
			\par\end{centering}
		
		\label{tab:vgg11}
	\end{table}
	
	\begin{table}[ht]
		\caption{Binary labels used in the proposed ordinal regression framework to predict the Bethesda score.}
		\centering
		
		\begin{tabular}{|c|c||c|c|c|c|}
			\cline{3-6} 
			\multicolumn{1}{c}{} & \multicolumn{1}{c|}{} & $S^0_l$ & $S^1_l$ & $S^2_l$ & $S^3_l$\tabularnewline
			\hline 
			\hline 
			& $S_l=2$ & $0$ & $0$ & $0$ & $0$\tabularnewline
			\cline{2-6} 
			Bethesda & $S_l=3$ & $1$ & $0$ & $0$ & $0$\tabularnewline
			\cline{2-6} 
			score & $S_l=4$ & $1$ & $1$ & $0$ & $0$\tabularnewline
			\cline{2-6} 
			& $S_l=5$ & $1$ & $1$ & $1$ & $0$\tabularnewline
			\cline{2-6} 
			& $S_l=6$ & $1$ & $1$ & $1$ & $1$\tabularnewline
			\hline 
		\end{tabular}
		
		\label{tab:ordinal_gr_truth}
	\end{table}
	

\end{document}